\documentclass[aps,
pra,epsfig]{revtex4}

\usepackage{epsf}
\usepackage{amsfonts}
\usepackage{amssymb}
\usepackage{graphicx}
\usepackage{color}
\usepackage{amsmath}
\usepackage[bookmarksnumbered, bookmarks, breaklinks, linktocpage]{hyperref}

\def\ket#1{| #1\rangle}

\def\R{\hbox{\rm I \kern-5pt R}}

\def\Tr{{\rm{Tr}}}

\begin{document}

\title{One world versus many: the inadequacy of Everettian accounts 
of evolution, probability, and scientific confirmation}

\author{Adrian Kent}
\email[]{a.p.a.kent@damtp.cam.ac.uk}
\affiliation{Centre for Quantum Computation, DAMTP, Centre for Mathematical
  Sciences, University of Cambridge, Wilberforce Road, Cambridge CB3 0WA,
  U.K.}
\affiliation{Perimeter Institute for Theoretical Physics,
31 Caroline Street North, Waterloo, Ontario, Canada N2L 2Y5}

\date{\today}

\begin{abstract}
There is a compelling intellectual case for exploring
whether purely unitary quantum theory defines a sensible and 
scientifically adequate theory, as Everett originally proposed.
Many different and incompatible attempts to define a coherent Everettian
quantum theory have been made over the past fifty years.  
However, no known version of the theory (unadorned by extra ad hoc postulates)
can account for the appearance of probabilities and explain why the 
theory it was meant to replace, Copenhagen quantum theory, 
appears to be confirmed, or more generally why our evolutionary history 
appears to be Born-rule typical. 

This article reviews some ingenious and interesting recent attempts in 
this direction by Wallace, Greaves-Myrvold and others, and explains
why they don't work.  An account of one-world randomness, 
which appears scientifically satisfactory, and has no many-worlds
analogue, is proposed.  A fundamental obstacle to confirming many-worlds
theories is illustrated by considering some toy 
many-worlds models.  These models show that branch weights can exist without
having any role in either rational decision-making or theory confirmation,
and also that the latter two roles are logically separate. 

Wallace's proposed decision theoretic axioms for rational agents in 
a multiverse and claimed derivation of the Born rule 
are examined.  It is argued that
Wallace's strategy of axiomatizing a mathematically precise decision theory 
within a fuzzy Everettian quasiclassical ontology is incoherent.   
Moreover, Wallace's axioms 
are not constitutive of rationality either in Everettian quantum theory
or in theories in which branchings and branch weights are precisely
defined.  In both cases, there exist coherent rational strategies 
that violate some of the axioms. 
\end{abstract}
\maketitle

\section{Introduction}

\subsection{Some common ground}

Although I disagree with the
Everettian contributors to this volume on some fundamental questions, 
I think they deserve much credit for developing some creative
and interesting ideas and arguments, which have
certainly helped advance our understanding of fundamental science. 
To elaborate on this, let me note some points on which I
agree with many Everettians.   

First, the Everettian programme had a sensible motivation.  Everett
asked\cite{everettone}, in effect, whether 
quantum theory really needs to be framed
in such a way that the evolution of the wave function is governed
by two different laws: generic unitary evolution together with the
projection postulate when measurement takes place.  
It's a good question.  Even if, rather than the projection postulate, 
quantum theory came equipped with a precise extra dynamical law
implying the postulate as an approximation, it would be natural to ask
if we really needed it.  As it is, there 
is quite a compelling case for exploring
whether we can make sense of purely unitary quantum theory.

Second, it {\it is} a sensible project to try to extract a physical
ontology from a unitarily evolving quantum state vector, given a
theory of the initial state or initial conditions, a Hilbert space
defining a representation of position, momentum and other canonical
operators, and a dynamical theory that expresses the Hamiltonian in
terms of these operators.  Whether the project succeeds
in producing an ontology with the properties that Everettians
fondly imagine is another question -- but certainly one worth discussing. 
One still, strangely, sometimes hears the argument that it is
illegitimate -- a basic misunderstanding of quantum theory -- even to
examine the possibility of giving the state vector a direct physical
interpretation.  This seems to me simply unimaginative dogma.
Everettians are right to insist that their programme 
should be judged on whether or not it works, not on whether it respects 
pre-Everettian quantum orthodoxies.

Third, neither the apparently fantastic nature of the Everettian
worldview, nor the superficial conflict between postulating multiple
independent mutually inaccessible worlds and Occam's razor, are
entirely compelling arguments against the Everett programme.  One
needs to consider Everettian ideas in the context of other attempts to
make sense of quantum theory, and in detail.  
One of the great intellectual challenges of theoretical physics is 
to find a mathematically elegant, universally applicable, 
Lorentz covariant, scientifically adequate version of quantum
theory that supplies a well-defined realist ontology.  
If the Everett programme really could produce a
well-defined Lorentz covariant physical ontology that adds little or
no arbitrary structure to the mathematics of quantum theory, and that 
reproduces all the scientific successes of Copenhagen quantum theory
within its domain of validity, it would have solved this fundamental
problem.  Given the present alternatives, we would, I
think, at that point, have to consider it seriously as a possible
account of reality.

Now, in fact, I think that the Everett programme fails
in these ambitions, for reasons explained below.  I am also 
optimistic\cite{akoneworld}
that we can find simpler one-world versions of quantum theory that have 
all the aforementioned virtues and none of the problems that afflict,
and I think ultimately doom, the Everett programme.  But I see no way
to make either conclusion so transparently true as to eliminate the
need for argument.

Fourth, it matters -- it is scientifically important to understand --
whether the Everettian programme can possibly succeed.  If Everettians
really could produce a theory of reality with all the proclaimed
virtues, it would clearly weaken (though not eliminate) the motivation
for other attempts at solving the quantum reality problem --
just as finding a consistent quantum theory of gravity would weaken
(though not eliminate) the motivation for looking for others.
Conversely, if, as I argue, the Everettian programme has fairly
definitely failed, then the problem of finding a viable formulation of
quantum theory applicable to closed quantum systems looms rather large
among the concerns of theoretical physics.  The failure of the Everett
programme adds to the likelihood that the fundamental problem is not
our inability to interpret quantum theory correctly but rather a
limitation of quantum theory itself.  If so, my guess is that
we most likely won't find
an adequate cosmological theory so long as we assume that quantum
theory is universally valid -- so we should be looking for possible
signals of the failure of quantum theory applied to the universe.
Likewise, if so, quantum interference quite likely breaks down
somewhere between the microscopic and the macroscopic -- so we should
be working harder to characterize the most promising types of
experiment to test this.

Everettian ideas have been around for fifty years, and influential for
at least the last thirty.  Yet there has never been a consensus among
theoretical physicists either that an Everettian account of quantum
theory can be made precise and made to work, or that the Everettian
programme has been comprehensively refuted.  These questions are quite
central to the future of theoretical, experimental and observational
physics.  We need to resolve them and move forward.

\subsection{Everett's elusive essence}

\begin{quotation}
``When he died, his heirs found nothing save chaotic manuscripts. His
family, as you may be aware, wished to condemn them to the fire; but 
his executor -- a Taoist or Buddhist monk --  insisted on their
publication.''

``We descendants of Ts'ui P\^{e}n,'' I  replied, ``continue to curse that
monk. Their publication was senseless. The book is an indeterminate
heap of contradictory drafts.''

(Jorge Luis Borges, {\it The Garden of Forking Paths} \cite{borges}) 
\end{quotation}

\begin{quotation}
`` $\ldots$ so crowded with $\ldots$ empty
sophistication that it is extremely difficult to
perceive the simple errors at the basis. It is like fighting the hydra --
cut off one ugly head, and eight formalizations take its place.'' 

(P.K. Feyerabend, {\it  How to Defend Society Against 
Science} \cite{feyerabend})
\end{quotation}

After fifty years, there is no well-defined, generally agreed set of
assumptions and postulates that together constitute ``the Everett
interpretation of quantum theory''.  Far from it: Everett\cite{everettone,everetttwo}, 
DeWitt\cite{dewitt}, Graham\cite{graham}, Hartle\cite{hartlefreq}, Geroch\cite{geroch}, 
Deutsch\cite{deutschone}, Deutsch\cite{deutschtwo}, 
Saunders\cite{saundersvol}, Barbour\cite{barbour} (partly inspired
by Bell\cite{bellmw}, though Bell's aim was not to inspire), 
Albert-Loewer\cite{almanyminds}, Coleman\cite{coleman}, 
Lockwood\cite{lockwood}, Wallace\cite{wallacevolone}, 
Wallace\cite{wallacevoltwo}, Vaidman\cite{vaidman}, 
Papineau\cite{papineauvol}, Greaves\cite{greaves}, 
Greaves-Myrvold\cite{greavesmyrvoldvol},
Gell-Mann and Hartle\cite{hartlevol}, Zurek\cite{zurekvol} and
Tegmark\cite{tegmarkvol}, among many others, have offered
distinctive and often fundamentally conflicting views on what 
precisely one needs to assume in order to get the Everett programme off
the ground, and what precisely an Everettian (or, some say,
post-Everettian) version of quantum theory entails.  

I am primarily interested here in contrasting realist ``one-world'' and
``many-worlds'' accounts of quantum theory.  By {\it one-worlders}, I mean
those who aim to find a version of quantum theory in which quantum
experiments really have only one outcome, we really have only one
version of our future selves at any future time, and some intrinsic
randomness in nature determines which outcome occurs and which future
self is realised from among the range of possibilities defined by the
theory.  For example, within its domain of validity, the Copenhagen
interpretation of quantum theory is a one-world theory.  By
{\it many-worlders}, I mean those who share Everett's view that a unitarily
evolving quantum state vector should be interpreted as directly
representing reality, and the future versions of ourselves that
observe different outcomes of quantum experiments should be
interpreted as equally real.

So, I will not discuss here attempts at ``post-Everettian''
interpretations like those of Gell-Mann and Hartle\cite{hartlevol} and
Zurek\cite{zurekvol}, which fall into neither camp, and 
seem -- despite much critical probing -- unclear on, or uncommitted to 
taking a stance on, precisely what, if
anything, in the theory corresponds to objective external reality.
(Extensive critiques of Gell-Mann and Hartle's approach
can however be found 
elsewhere\cite{dowkerkentone, dowkerkenttwo, 
kentone, kenttwo, kentthree, kentfour}.)

My main focus is on the recent attempts by 
Wallace\cite{wallacevolone, wallacevoltwo}, Greaves and 
Myrvold\cite{greavesmyrvoldvol}, and, to a lesser extent, 
Papineau\cite{papineauvol} and Saunders\cite{saundersvol}, 
to define, analyse and test realist many-worlds 
interpretations.  These authors offer
different, and on some points mutually inconsistent, approaches, but
nonetheless share enough common perspectives to be considered
together.  Their papers include some very interesting and creative
arguments, which raise important scientific questions.  However, I
will argue below that none of their approaches produces a
scientifically adequate version of quantum theory.

Shadowing these discussions is the spectre of the 'many-minds
interpretation' set out some time ago by Albert and Loewer
\cite{almanyminds}.  Essentially everyone, including Albert and
Loewer, agrees that the many-minds interpretation, while logically
consistent and in accord with the data, is utterly unsatisfactory,
since it adds to the Everettian formalism a collection of ad hoc
postulates which not only are (even by Everettian standards)
fantastic, but also undercut the motivation for taking Everett
seriously, namely that it purports to explain how to make sense of
quantum theory without adding extra equations or interpretational
postulates.  So, no one -- certainly no one represented in this 
volume -- wants to be a many-minder.  And here lies the problem: it seems to
me (and to others -- see in particular Albert's 
contribution\cite{albertvol} to this
volume) that, at various points in their arguments, Saunders, Wallace,
Greaves-Myrvold and Papineau tacitly -- and, since they reject the many-minds
interpretation, illegitimately -- appeal to many-minds intuitions.
Indeed, at least in the first three cases, it seems to me that if one
fleshed their ideas out into a fully coherent and complete
interpretation, one would necessarily arrive either at the many-minds
interpretation or something even worse.  I will elaborate on this
below.
   
Of course, these discussions crucially turn on our understanding of what
counts as scientifically adequate.  The idea that reality contains
many essentially independent quasiclassical worlds corresponding to
different possible cosmological and experimental outcomes clearly
isn't, per se, susceptible to logical refutation.  That isn't at
issue.  The key question, to my mind --- and I think
modern Everettians, including the authors considered here,
generally agree --- is whether we can find an appealingly simple version of
quantum theory in which a realist many-worlds ontology is essential
(i.e. there is no equally simple one-world variant) and which (at
minimum) replicates all the scientific successes of one-world quantum
theory (i.e. quantum theory including some form of the projection
postulate, or some principle from which it can, approximately and
within a suitable domain of validity, be derived).  I believe that we
can't.  In particular, it seems to me the
Everettian programme has not produced and cannot produce a scientifically
adequate alternative account that reproduces the standard one-world
account of probabilistic inferences derived from quantum theory
-- despite the ingenious recent attempts of
contributors to this volume.  

Some commentators sympathetic to the Everettian programme\cite{greavesmyrvoldvol, papineauvol} 
argue that a double standard is at work here: that criticisms
of the Everettian programme's attempt to account for the appearance
of probability can and should equally well be applied to the
standard understanding of the role of probability in one-world
versions of quantum theory, and indeed of probabilistic scientific
statements in general.   To respond to this point, I consider below
some fundamental differences between 
randomness (or apparent randomness) in one-world quantum theory and 
its purported Everettian analogue, and point out what seem to me
irresolvable problems with the latter.

\section{One-world theories and probability}\label{oneworldprob}

Copenhagen quantum theory is a one-world version of quantum theory:
any given experiment or quantum event has a number of possible
outcomes, but only one actual outcome.  Some other non-Everettian
variants and modifications of quantum theory, such as de Broglie-Bohm
theory and dynamical collapse models, similarly randomly select from
many possible physical evolutions, and can be (and usually are)
interpreted as defining a unique quasiclassical world.  The consistent
histories approach \cite{hartlevol}, if combined with 
an (alas unknown) suitable set
selection rule, would also lead naturally to a one-world
interpretation, in which reality is described by one randomly chosen
history from the selected set.  And these by now venerable contenders
certainly don't exhaust the possible options.\cite{akoneworld}
My aim here is not to advocate a specific one-world version or variant
of quantum theory, or to assess the current state of the art, but
rather to compare and contrast one-world and many-world accounts of
probability.  For that purpose, let us suppose, for the sake of
argument, that we have to hand a particular one-world theory that 
implies that, while the universe could have evolved in a (presumably
very large) number of different ways, one quasiclassically evolving
world -- the one we observe -- was randomly selected.

One-world versions of quantum theory, together with hypotheses about
the initial conditions and unitary evolution, predict the
probabilities of our experimental results and observations.  We test
the theory and these hypotheses by checking whether the results are of
a form we would typically expect given the predicted probabilities.
In practice, pretty much everyone agrees on the methodology of theory
confirmation, at least sufficiently so that, for example, everyone
agrees that, within the domain of validity of Copenhagen quantum
theory, the Born rule is very well confirmed statistically.  However,
there is much less agreement on how, or even whether, we can make
sense of fundamentally probabilistic physical theories.  What exactly, if
anything, does it mean to say that the probability of the universe
turning out the way it did was $0.00038$?

Everettian authors have stressed this last point lately.  We should
not, they argue, apply different standards to one-world and
many-worlds quantum theory.  If our account of standard probability
applied to one-world quantum theory is suspect, or incomplete, or
involves ad hoc postulates, we cannot reasonably reject an alternative
many-worlds account on the grounds that it runs into difficulties
that might, on close analysis, turn out to be precisely analogous.

There are several possible responses for one-worlders here.  One
response is to try to defend or buttress or further develop
frequentism, or another standard account of standard probability.  A
second is to try to point out some insuperable problems with many-worlds
accounts of probability, and thus make the case that, whatever
difficulties one-world quantum theory might run into, many-worlds
quantum theory cannot possibly be satisfactory.  A third is to argue
that the difficulties that many-worlders face in dealing with
probability are worse than -- not, as claimed, precisely 
analogous to -- those faced by one-worlders.

I think the first of these options is worth pursuing.  I think too
that the second and third lines of argument are valid, and I will
develop them later.  But, in this section, I want to make a separate
point.  I want to suggest a non-standard account in which 
the scientific space usually occupied by
one-world probabilistic theories is filled instead by 
deterministic theories with
a large amount of theoretically unspecified data.  This allows us to
compare, verify and falsify theories, and to recover essentially all
of current science, without assigning a fundamental role to
probability {\it per se}.   Convinced believers in a chancy world might regard
this as a useful fall-back position pending a fully satisfactory 
explanation of standard probability.   It might, alternatively, 
be seen as an account with enough attractions of
its own that it could be preferable to
any standard account involving probability.  Either way, it offers 
a way of making scientific sense of one-world quantum theory that 
has no many-worlds analogue. 

Consider a probabilistic theory $T$, and suppose for simplicity that
it predicts a finite set of probabilistic events, labelled by the
index $i$, each with finitely many possible outcomes $x^j_i$, labelled
by the index $j \in J_i$, for which it predicts nonzero probabilities
$p^j_i$.  For simplicity, we also suppose for the moment that the
possible outcomes for any given event, and their probabilities, are
independent of the outcome of any other event.  We say two events $i$
and $i'$ are of the same type, according to $T$, if the sets $\{p^j_i
\}$ and $\{ p^j_{i'} \}$ are identical.  Let $B = \{ 0 , 1 \}$, $B^* =
\{ \emptyset , 0 , 1 , 00, 01 , \ldots \}$ be the set of finite binary
strings, and $B^r$ the set of length $r$ binary strings.  Let $n =
\prod_i | J_i | $ be the size of the list of possible sets of
outcomes, which we write as $N = \{ 1 , 2 , \ldots , n \}$.

A length $r$ code for the outcomes is any surjective map $C: B^r
\rightarrow N $.  Given such a code, we can define an alternative
probabilistic theory $T^C$ by stipulating that a binary string $b$ in
$B^r$ is randomly chosen from the uniform distribution, and that the
outcomes are given by $C(b)$.  By taking $r$ sufficiently large, and
choosing $C$ so that $ | \{ b : C(b) = i \} | \approx 2^r p(i ) $ for
each $i \in N$, we can find theories $T^C$ whose probability
assignments are arbitrarily close to those of $T$.

A length $r$ subcode for the outcomes is any map (not necessarily
surjective) $C: B^r \rightarrow N$.  Again, given a subcode, we can
define a probabilistic theory $T^C$ as above: here $T^C$ may assign
zero probability to some outcomes for which $T$ assigns non-zero
probability.

We can define another type of theory from the triple $( T^C , C , r)$:
a theory that simply states that the data will be those predicted by
$T^C$ and $C$ given some length $r$ binary string as input, and makes no
prediction about the binary string.  We call this theory $D(
T^C , C , r)$, using $D$ to emphasize that we now regard the theory as
deterministic.  One might view the binary string in $D ( T^C , C , r
)$ as playing a role analogous to a constant of
nature in a deterministic theory: its value is not
fixed by the theory, and can only be determined empirically. 
In this case, even if the map $C$ is injective,
determining the entire string would require observing 
every random event in the universe.

Now, on the view that there is a unique ``correct'' fundamentally
probabilistic theory of nature $T$, each probabilistic theory of
the form $ T^C $ must be either equivalent to $T$ (which is
possible only if the probabilities $p^j_i$ are all dyadic), or else
incorrect (though possibly a good approximation to $T$).  Note though
that, given a finite set of data, many other probabilistic theories
besides $T$, including some of the form $T^C$, will be consistent with the data.
Indeed, we would generally expect some theories $T'$ to fit the data
better than $T$, in the sense that the same sets of events are of the
same type according to $T$ and $T'$, and the probabilities $p'^j_i$
are closer than $p^j_i$ to the observed relative frequencies for
events of the same type.  If we nonetheless regard $T$ as likelier to
be correct than $T'$, it must be for reasons other than purely
empirical -- presumably on grounds of elegance or simplicity.  
And if we maintain that there is a unique correct fundamental theory, 
it seems to follow that the correct theory is determined by a
set of probabilities $\{ p^j_i \}$ not determined by 
the physical universe (although perhaps very well approximated by relative
frequencies of physical events).  

Here's an alternative view.   It may be, if not
meaningless, then at least unnecessary, to appeal to the idea of a
unique correct fundamentally probabilistic theory of nature, or even
to define probability as a fundamental physical concept.  Instead of
considering probabilistic theories $T^C$, we can compare deterministic
theories $D ( T^C , C , r )$ against one another and against the data.
In evaluating these theories, we use the criteria of simplicity and
elegance.  These criteria have no precise mathematical definition. 
They include judgements about the form of $T^C$ and $C$, as well as
the parameter $r$ (which is a precise measure of complexity for the
part of the theory defined by the unknown binary string).  In saying
that one theory $D( T^C , C , r )$ is our best current theory -- or
perhaps that our best descriptions of nature are given by a
class of similar such theories -- we mean that we can't presently find
a substantially simpler and more elegant theory that fits the data.  
The stronger meta-theoretic hypothesis that a theory $D( T^C , C , r ) $ is, up to 
approximate equivalence, {\it the} best theory of nature implies 
that, given all the physical data in the universe, one would not be able to
find a simpler, more elegant, compelling theory.

This could be made more quantitative by formalising the
discussion within the context of a fixed model of computation, for
instance a (classical) Turing machine.  (This is why we have chosen to
consider theories with unknown binary strings, although of course
bases other than binary could also be used.)  Here, a 
theory is a program for generating a mathematical representation of
the complete set of physical data.  A theory with unknown data is a
program that requires an unknown input string of stipulated length.
The theory's simplicity depends, inter alia, on both the length of the
program and the length of the required input string.  Each of these
is a natural simplicity parameter.  The halting time
of the program is another significant parameter, which gives one way 
of quantifying the elegance of a theory.  

In principle, within a fixed computation model, 
it's possible to carry out an exhaustive search of all
theories with total length $ \leq L$ that halt after $\leq N$ steps.
In principle, thus, given all the physical data, one can test the
hypothesis that $D ( T^C, C, r)$ is the best theory among all those whose 
program and input strings satisfy given length bounds, and which
satisfy other stipulated simplicity and elegance constraints,
that halt after any given finite time, relative to a 
fixed computation model.  

Thus, instead of talking
about a probabilistic physical theory that produces a random set of
physical data, we can consider a deterministic physical theory whose
definition includes a set of pre-determined but a priori unknown physical data, 
together with the meta-theoretic
hypothesis that this description is essentially algorithmically 
incompressible.  If we learn empirically that the data are in fact
significantly compressible, then this hypothesis is refuted, and we
may replace the theory by a more economical one.  

It should be stressed that these measures of simplicity and 
elegance are by no means intended to be an exhaustive list. 
For example, another elegance criterion is given by the 
principle of scientific induction, which suggests we should prefer a theory
that suggests that a hitherto apparently fair coin will continue to be 
apparently fair over one that suggests that it will henceforth always
come up heads, even though the latter theory requires a shorter input 
string (and so is simpler by one of the above 
measures).\footnote{I thank Jonathan
Barrett for this point and this example.}  
Comparing scientific theories generally involves 
a wide and arguable variety of quantitative and qualitative simplicity and 
elegance criteria, and nothing in this account alters that: the aim here 
is only to propose a different treatment of apparent randomness when
comparing theories.  

\subsection{Example: reinterpreting a fair coin}

For example, in a universe with an apparently
random process that apparently mimics a fair coin and produces a large
number $N$ of apparently independent outcomes, 
our meta-theoretic hypothesis might suggest that we cannot find a
simpler correct theory than one that states that
the length $N$ binary string is essentially algorithmically incompressible.
If, in fact, the string $S$ turns out to consist of $0.01 N$ zeroes and $
0.99 N$ ones, we can certainly generate a more economical
theory, and this hypothesis is refuted.

According to the standard account of probabilistic theories, if a
probabilistic theory $PT$ says that zeroes and ones are
equiprobable and independently generated, the outcome $S$ is 
extremely improbable, but not logically impossible.  The theory $PT$
is thus not logically refuted by the outcome $S$.  In 
practice we would reject it -- but, without a fundamentally
satisfactory account of probability, it is hard to give a completely
satisfactory justification for doing so.    
 
In our alternative account, however, no such problem arises.  
Our hypothesis predicts that a
given physical dataset is essentially incompressible --- where 
``essentially'' incorporates some judgements about the tradeoffs between
small gains in compression of the dataset and simplicity and
elegance in other aspects of the theory.  If the dataset turns out to
be a string such as $S$ that is 
significantly compressible, so that we can fit the
data by a simpler theory, the hypothesis is falsified and the original
theory replaced.

\subsection{Example: reinterpreting a biased coin}

Now consider a universe with an apparently
random process that apparently mimics a coin with bias $p > \frac{1}{2}$ 
towards zero and produces $N$ apparently independent outcomes.   
We can then produce theories that state that 
the length $N$ binary string is compressible to $H(p) N + o (N)$ bits.  
For example, a theory which says that the length $N$
string will contain between $p N - 10 \sqrt{N}$ and 
$p N + 10 \sqrt{N}$ zeroes has
the required compression, since we can binary code all such 
strings in a code of length $H(p) N + o (N)$.   
Clearly there are many somewhat similar such theories --- the 
string contains between
$p N - 9 \sqrt{N}$ and $p N + 9 \sqrt{N}$ zeroes, between $p N - 11 \sqrt{N}$ and
$p N + 11 \sqrt{N}$ zeroes, and so on. 
On this view of scientific accounts of apparently random data, that's 
the best one can hope for: generically, no single clearly optimal theory 
will emerge.   However, we can hypothesize that theories of roughly 
this length are essentially best possible -- i.e. that the string 
cannot be compressed to significantly shorter than $H(p)N$ bits --
and {\it this} hypothesis is testable and falsifiable.   

Again, these theories reproduce deterministically predictions
that the standard probabilistic theory says hold with probability
very close (but not equal) to one.  They exclude some very low probability events
which would, if realised, in practice persuade almost everyone
that the probabilistic theory was wrong, even though their occurrence
is logically consistent with the theory. 

\subsection{Conclusion}

According to this account, we should consider one-world quantum
theory as a theory which requires a binary string as input, 
and consider it alongside the meta-theoretic hypotheses that (a) there is no 
significantly more compressed
description of the data obtained from quantum experiments than that
given by encoding them in binary, using a coding that would produce an
approximately uniform distribution over binary strings if the data
were probabilistically generated via the Born rule, (b) the data 
can indeed be thus described.  If one of these hypotheses
turns out to be incorrect -- if, for example, the data in all Bell experiments
consistently show significantly greater violations of the CHSH
inequality than quantum theory predicts -- then we must find
a better theory.  Conversely, the theory logically (not merely with high
probability) implies that we will see no consistent regularities in
our experimental data that would, on the usual account, be highly
improbable.

Among the scientific virtues of this account, as I see it, are its
explicitness about the provisional nature of our theories, and its
undogmatic sidestepping of the problem of giving a fundamental
meaning to probability.  It recognizes
the possibility that random-seeming data may turn out to have a
simpler description.  It recognizes too that, if we find consistent
regularities that a probabilistic theory says are highly improbable,
then we should and will feel impelled to produce a better theory.  At
the same time, it stays silent on the question of whether
random-seeming physical data are genuinely randomly generated in some
fundamental sense, and hence avoids the need to explain what such an
assertion could really mean and how we could be persuaded of its
truth.

One-world quantum theory, read 
in this way, allows us to draw logical inferences about the physical
world.  It predicts -- it is not merely consistent with the fact --
that there will be no regularities in the data of a type that would
allow for a significantly simpler theoretical description.  If that
prediction turns out to be wrong, the theory is refuted.  Interpreted
thus, one-world quantum theory can be read as a well-formulated
scientific theory, in a way that allows a straightforward account of
scientific confirmation and refutation.  If we assume it is correct,
we have an explanation for the apparent fact that our evolutionary and
experimental histories contain no regularities that would be
inexplicably improbable according to the Born rule.  
To the extent that the project outlined above can be fleshed out
and succeeds -- and I am optimistic that it can and does -- 
proponents of one-world quantum theory can rest relatively
easy on the question of randomness.  

\section{Toy many-worlds theories and their uses}

If we knew of probability theory 
{\it only} through its use in Copenhagen quantum theory -- if we had 
no familiarity with coin tosses, dice rolls, noise, or any other 
effectively unpredictable classical systems -- we would probably
be (even more) deeply confused about the nature of both quantum theory 
and probability.   I suspect this is the cause of much of the continuing
confusion over 
many-worlds quantum theory: discussions need simultaneously to grapple
with the quite unfamiliar concept of many branching worlds and the 
specific peculiarities of Everettian quantum theory.   

This motivates defining some simpler many-worlds theories.   Another 
reason for doing so is that 
some key Everettian ideas -- for example, Greaves and Myrvold's 
attempt \cite{greavesmyrvoldvol} at an account of many-worlds theory confirmation -- can really only 
sensibly be discussed if we can consider a class of many-worlds
theories, not just the single example of Everettian quantum theory.   
Readers may initially find the form of the 
following theories a little intellectually unsettling, 
but I recommend persevering: they shed a great deal of 
light on Everettian arguments.  

Let me stress right away that these are not perfect models for 
Everettian quantum theory.  That is, in fact, part of the point:
they allow us to separate out general claims about rationality and
theory confirmation in multiverse theories from claims that
rely on specific features of quantum theory. 
In particular, they allow us to see why Greaves-Myrvold's 
account of many-world theory confirmation doesn't work.    

\subsection{Some toy multiverses}

The following toy multiverses are all classical, in the sense that the 
state of any branch at any time is defined by a classical physical theory,
and they all have a definite branching structure.   

Consider, first, the branching multiverse
$CBU_1$, which includes 
conscious inhabitants, and also includes a machine with a red button 
on it and a tape emerging from it, with a sequence of numbers on it,
all in the range $0$ to $(N-1)$.   Whenever the red button is pressed in 
some universe within the multiverse, that
universe is deleted, and $N$ successor universes are then created. 
All the successors are in the same classical state as the original
(and so, by hypothesis, all include conscious inhabitants with the
same memories as those who have just been deleted), except that 
a new number has been written onto the end of the tape, with the 
number $i$ being written in the $i$-th successor universe.   

Suppose, too, that the multiverse's inhabitants believe 
that something like this is indeed happening.   The numbers on the 
tape play a significant role in their society.  
In particular, it is quite common to place bets on future numbers, and 
social mores ensure that such bets are always honoured.  
Of course, since one's own universe will be destroyed before the
next number is written, placing such a bet means -- they correctly
believe -- redistributing resources amongst one's successors.   
Some inhabitants may find reasons for preferring some redistributions 
over others.  We need not discuss yet precisely what these reasons and 
preferences (both of which may be different for different
inhabitants) may be.  

It might be helpful to imagine that
the universes are being run on a simulator by technologically
advanced beings, who simply end one simulation whenever the red
button is pressed, and then start simulating the successor universes
from the appropriate initial states.  We will sometimes assume 
that the inhabitants, indeed, believe this to be the case.  

Suppose, further, that some of the inhabitants of $CBU_1$ 
have acquired the 
theoretical idea that the laws of their multiverse might attach
{\it weights} to branches, i.e. a number $p_i$ is attached
to branch $i$, where $p_i \geq 0$ and $\sum_i p_i =1$.
They may have different theories about how these weights are
defined: for instance, that the weights are always $\{ p_i \}$,
that they are always $\{ q_i \}$, that they vary over time 
according to some rule, and so on.   As it happens, though, these
theories are all incorrect: there are no weights attached to the branches.
To be clear: this is not to say that the branches have equal weight.
Nor are they necessarily physically identical aside from the tape numbers. 
They may perhaps be distinguished by other features: for example, if 
they are simulations, they may be simulated by different hardware or software.   
However, any such differences do not yield any natural quantitative 
definition of branch weights. 
There is just no fact of the matter about branch weights in this multiverse.

The multiverse $CBU_2$
is similar to $CBU_1$.   In this universe,
there {\it are} indeed numbers attached to the branches, but the way 
they are attached means that they should (by our lights, and also
by the inhabitants', if only they understood the full picture)
have no significance to any 
decisions the inhabitants make about bets/redistributions.  
For instance, we could extend the simulation idea, and imagine
that the technologically advanced beings simply choose, on whim,
to write the number $p_i$ somewhere inconspicuous in the simulation 
of successor universe $i$, in such a way that it has no effect 
on the inhabitants, and that it has no other significance.   

The multiverse $CBU_3$ is similar to $CBU_2$.  However, this time
the numbers attached to the branches by the physical theory are
attached in such a way that it can be plausibly argued that 
they {\it could} reasonably play a significant role in the decisions the
inhabitants make about bets/redistributions.  
For instance, we could imagine that when the technologically advanced beings 
create successor universes, they create not just one successor corresponding
to each outcome $i$, but a number of distinct successor universes, all identical
apart from their outcome values, and the number containing outcome $i$ 
is proportional to the weight $p_i$.   (We assume here
the $p_i$ are rational numbers.)  

\subsection{Some possible strategies}\label{strategies}

Consider an inhabitant of any of the above multiverses,
who believes that the weight $p_i $ is attached 
to the outcome $i$.    
Suppose they are offered a variety of bets
that give their successor a good $G_i$ in
a universe in which outcome $i$ obtains, and they (the original
inhabitant) attach utility $U_i$ to this good.  We suppose the $U_i$ 
are finite real numbers, not necessarily positive (the goods may be
bads); and, of course, both $G_i$ and $U_i$ depend on the bet. 

How might they proceed to evaluate and rank such bets?  
{\it Weight-sensitive} inhabitants believe
that branch weights exist and should play a role in their 
betting preferences.  
{\it Weight-indifferent} inhabitants may also 
believe that the physical theory 
attaches weights, but if so, do not believe  
they are of any relevance to a rational betting strategy.  (Such an
inhabitant might, for example, believe that they live in 
a multiverse like $CBU_2$.)
Among their options is to mimic the strategy of a weight-sensitive
inhabitant, except that they treat all branch weights as equal.  
By this means, given any weight-sensitive strategy, we can define
a corresponding weight-indifferent strategy. 
Here are some examples of weight-sensitive strategies: 

\begin{itemize} 

\item The {\it mean utilitarian} ranks bets according to the value 
of $\sum_i p_i U_i$.   

\item The {\it Price-Rawlsian}'s dominant concern\cite{pricevol,rawls} 
is with the welfare
of their least satisfied future self.   They rank bets first according to 
$\min  ( U_i )$, and then some list of tie-breaking criteria.   
To be definite, let's say their next criterion is the value of 
$ \sum{p_j} $, summed over all $j$ such that $U_j = \min ( U_i )$, 
followed by $\min ( U_j: U_j \neq \min (U_i ) )$, and so on. 

\item The {\it future self elitist}'s dominant concern is that the best 
possible version of their future self should be realized 
somewhere; they have little
interest in mediocre future selves, whom they regard as losers.  
Their bet rankings are thus dominated by $\max ( U_i )$, and they break
ties using the mirror image of the Price-Rawlsian's criteria.  

\item The {\it rivalrous future self elitist} takes things one stage
further.  Not only do they identify their interests exclusively with
those of their best possible future self, but they regard that self as in
competition with the others, and feel happiest -- all else being
equal -- if that competition is won by as large a margin as possible.
They rank bets first by $\max (U_i )$, then by $ \max ( U_i ) - 
\max ( U_j : U_j \neq \max (U_i ) ))$, and so on.  

\item The {\it median utilitarian}'s dominant concern is for 
median utility.  Reordering the index labels so that 
$U_1 \leq U_2 \leq \ldots \leq U_n$, let $j$ be such that 
$\sum_{i=1}^{j-1} p_i < \frac{1}{2}$ and 
$\sum_{i=1}^{j} p_i \geq \frac{1}{2}$: they rank bets first according
to the value of $U_j$.  
(They also have some tie-breaking criteria: one option is to
break ties by considering the mean utility.)  

\item The {\it $x$-percentile utilitarian}'s dominant concern is for
the utility of the future self ranked at $x$\% in the distribution.
They proceed like the median utilitarian, with $\frac{1}{2}$ replaced
by $\frac{x}{100}$.    The Price-Rawlsian, median utilitarian
and future self elitist are all special cases. 

\item The {\it future self democrat} believes her preference should
be that which would result from a democratic vote among her future
selves.   Given a finite list of possible bets, for each value of 
$x$, she asks herself how she would order her preferences 
among the bets, if she knew that she would become the future self ranked at 
the $x$-percentile of the elected bet.  
(The answer might be that her future self's voting preference would always be 
dominated by its own welfare under this hypothesis, but it need not: 
it depends whether she cares about the welfare 
of contemporaneous versions of herself in other branches.) 
She then tallies the votes, integrating over $x$ using branch weight
measure, and using, say, a single transferable vote system.  
The winner of the vote
is her preferred bet.   If the election is tied, she has  
more than one equally preferred bet.   

\item An example of a {\it future self distribution engineer} 
is someone who seeks to maximise an expression of the form
\begin{equation}
\sum_i f_1 (U_i ) p_i + \sum_{ij} f_2 (U_i , U_j ) p_i p_j + \ldots \, , 
\end{equation}
where the $U_i$ are the utilities of future branches with weights $p_i$, 
and $f_n$ is some given joint function of $n$ 
variables.\footnote{There are more general possibilities.}

\end{itemize}

\subsection{Many-worlds rationality reconsidered in toy models} 

\begin{quotation}
``If you do what you've always done, you'll get what you always got.''

(variously attributed)
\end{quotation}

According to Wallace \cite{wallacevoltwo} and 
Greaves-Myrvold \cite{greavesmyrvoldvol}, we should 
{\it define} rational behaviour in a multiverse via axioms 
generalizing those proposed by Savage\cite{savage} in order
to justify using the standard calculus of probabilities and
utility functions for rational decisions in a single world
in which future events are uncertain.   

\subsubsection{Savagean rationality in one world}

Savage, engagingly and rather admirably, 
presented his approach to rationality in 
the presence of (one-world) uncertainty 
\begin{quotation}
``$\ldots$ in a tentative spirit, for I realize that the serious blemishes
in it apparent to me are not the only ones that will be discovered by 
critical readers.'' \cite{savagequote} 
\end{quotation} 

Everettian neo-Savageans \cite{wallacevoltwo,greavesmyrvoldvol}, as I 
read them, 
seem rather less self-critical --- puzzlingly so, since applying 
Savagean decision 
theory to Everettian quantum theory raises many new questions without 
solving any
of the old ones.  This raises some general worries, which 
are developed to some extent
elsewhere in this paper, but might also be taken in other directions.      
First, if Savage's axioms are, in fact, unable to 
give a completely satisfactory account of 
ideal rational behaviour in the presence of one-world uncertainty, 
it seems very unlikely that a completely satisfactory axiomatic 
treatment of many-worlds rationality can be produced by generalizing them.  
Second, giving a satisfactory 
account of ideal rational behaviour in the presence of one-world 
uncertainty (or 
some many-worlds generalization) may in any case not be enough. 
(For one thing, we are not ideal rational agents.  For another, as 
Albert \cite{albertvol} has
eloquently stressed, there is a crucial difference between showing that 
one can find a rational justification for behaving as 
though the world were a certain way and 
showing that the world actually is that way.) 
Third, however far Savage can or cannot guide rational agents 
in one uncertain world, it isn't obvious
that his programme generalizes {\it at all} to many-worlds theories
in general or to Everettian quantum theory in particular.  

\subsubsection{Many-worlds rationality according to Greaves-Myrvold}  

Let me now focus
on Greaves-Myrvold's axioms, which are intended to apply to general
many-worlds theories, and so can straightforwardly be considered
within the toy models described above. 
I will consider later Wallace's 
arguments, which are framed for the special case of Everettian
quantum theory, and for the moment simply note that
their logic suggests the same conclusions here as Greaves-Myrvold's. 

In Greaves-Myrvold's view, the mean utilitarian's 
strategy is rationally justifiable, and
the others are branded irrational, since they violate one or
more of the axioms.  For example, 
the future self elitist and the Price-Rawlsian violate their
continuity postulate, $P6$, the median utilitarian violates
$P2$, the future self democrat violates
transitivity, $P1a$, and the rivalrous future self elitist
the dominance postulate $P3$.  

However, the fact is that each of these strategies
is well-defined and has a coherent motivation
(and many other such examples could also be constructed).  To brand them 
irrational seems to me itself irrational dogma.  Even the 
most contentious case, the rivalrous future self elitist, has a 
coherent, if ungenerous, philosophy of life in 
the multiverse and a rational strategy for implementing it.  
Note too that some of these  
strategies have arguable theoretical advantages over the mean utilitarian
strategy.   For instance, one can be an $x$-percentile utilitarian,
or a future self democrat (if they are {\it purely self-concerned}, in the 
sense that each future self's preferences among options are
completely determined by the implications for its own welfare), 
without having to quantify the utility
of the possible outcomes: one needs only a preference ordering.
This is arguably advantageous, since even if one accepts 
Savage's postulates \cite{savage} and, hence, the conclusion that one's 
preferences must be defined by some utility function, 
it may be difficult or even impractical to compute the 
relevant function for general outcomes, and yet relatively easy
to identify preferences among any finite list of outcomes. 

In short, Greaves-Myrvold's postulates only 
express in more abstract form a preference for being a mean utilitarian -- 
i.e., for one possible choice among many.   Their postulates are plausible
possible prescriptions for rational behaviour when considering the welfare 
of a population of future selves, but also logically inconsistent with
other plausible prescriptions.   This shouldn't come as a complete 
surprise: after all, Arrow's celebrated 
impossibility theorem \cite{arrow} taught us that 
plausible decision theoretic principles for populations may be inconsistent.  

Granted, the one-world counterparts of some of these strategies
may look peculiar.    But one can consistently accept the many-worlds
strategies as rational and reject their one-world 
counterparts.  As Price \cite{pricevol} has persuasively argued, many-worlds
agents can offer reasoned justifications for
their strategies that aren't available to their
one-world counterparts.  The many-worlds 
future self elitist knows that his best
possible future self will be an {\it actual} future self, while his one 
world counterpart doesn't.  The many-worlds future self 
democrat knows that there really will be a population of future selves 
who have preferences among the betting choices, while her one world
counterpart knows there won't be; and so on.

One could, of course, adopt a weaker position.
One could take Greaves-Myrvold's and Wallace's accounts 
of rationality as simply suggesting a possible attitude one {\it might} 
adopt to life in an Everettian multiverse, an attitude
defined by a set of rules which
are consistent and have some pleasant mathematical features but which
are not meant to constitute a dogma.   
On this liberal reading, Greaves-Myrvold's 
preferred strategy could be termed ``rational'', in the sense of 
being well-defined and internally consistent, without  
denying the existence of other equally rational
strategies.   The problem is that abandoning any claim of 
uniqueness also removes the purported connection between theoretical
reasoning and empirical data, and this is disastrous for the 
programme of attempting to interpret Everettian quantum theory
via decision theory.  If Wallace's arguments are read as
suggesting no more than that one can consistently adopt the Born rule
if one pleases, it remains a mystery as to how and why we arrived at
the Born rule empirically.   If Greaves-Myrvold's arguments are read 
as merely suggesting a possible attitude one might choose to take about 
testing and confirming many-worlds theories, one's left to investigate 
how many other equally valid attitudes there might be, and whether 
they mightn't -- disastrously -- imply
the confirmation of inconsistent theories from the same data.  

\subsection{Rationality and feasibility}

Consider now a rather more complicated multiverse, $CBU_4$.  Here, the 
universes are definitely being simulated by technologically advanced
beings, and the inhabitants know it.   They also know that, after the
red button is pressed, there is a list of outcomes $i$, and that the
list is indeterminately long (and possibly infinite).
They know too that, for each $i$, some number
of successor universes containing outcome $i$ will be created.   
They do not know
the number of successors there will be of each type: 
these vary for each $i$, and vary each time the 
button is pressed, at the whim of the simulators.   
What they do know -- because, let's say, the simulators have 
credibly promised them --
is that numbers playing the role of additive weights, following certain
rules, will be written inconspicuously into each simulated universe.   
Thus, if their universe has the number $x$ written in it, and the 
red button is pressed, and there are $n_i$
successor universes with outcome $i$, these successors
will have numbers of the form $ x q^j_i $ written into them, 
where the label $j$ runs from $1$ to $n_i$, $q^j_i \geq 0$, 
and 
\begin{equation}\label{weightrelation}
\sum_j q^j_i = p_i \, .
\end{equation}
   Here the $p_i$ are known to be constants (i.e. 
they take the same value each time the button is pressed), with $p_i \geq 0$,
and $ \sum_i p_i = 1$. The inhabitants know the values of a finite set of the $p_i$,
those with index $i \in I$, whose sum $ \sum_{i \in I} p_i < 1 $. 

What the inhabitants would {\it like} to do -- what they feel rationality
would mandate they do if they could -- is express betting/distribution
preferences that value each successor universe equally.   But they
can't -- they don't know how many successors will be created for any
given $i$, nor do they know how long the list of possible outcomes $i$ is.
Nor can they express betting/distribution preferences that value each
outcome equally, regardless of the number of successor universes containing
it -- again, they don't know how long the list of possible outcomes is.
  
What they {\it can} do is express betting/distribution preferences, for
bets on the known possible outcomes, treating the known values of $p_i$
as probability weights.   Doing so is equivalent to treating a successor
universe with the number $y$ written into it as having an importance  
proportional to $y$ -- a rule which can be consistently applied, despite
their ignorance about the number of successors of each type, because
of equation (\ref{weightrelation}).   So, they have a consistent,
feasible strategy available to them.   Moreover, if they want to 
assign a measure of importance to each individual universe, and they
want the importance they assign to the set of universes containing
outcome $i$ to be independent of the number of such universes, this 
is the {\it only available} rule.   Nonetheless, it doesn't
seem to have a fundamental rational justification.  The numbers 
written into the universes happen to follow convenient bookkeeping
rules, but they have no significance: there is no fundamental
{\it reason} to treat the numbers as a measure of importance of their
universes. 

From this, I think we should conclude two things, to be borne in
mind when we come to consider Wallace's arguments.   It can make perfect
sense, in a multiverse theory, to say that there exists a rational optimal
strategy that is inaccessible to the agents in that multiverse.   
Conversely, the fact that a strategy is available does not 
{\it per se} make it rationally compelling, even if it is the unique
available strategy satisfying some pleasant consistency properties: 
rational compulsion also needs rational justification, which 
may or may not exist.   

\section{Why many-worlds theory confirmation doesn't work}

Everettian quantum theory is essentially useless, as a scientific theory, unless
it can explain the data that confirm the validity of Copenhagen
quantum theory within its domain -- unless, for example, it can 
explain why we should expect to observe the Born rule to have
been very well confirmed statistically.   Evidently, Everettians cannot 
give an explanation that says that all observers in the multiverse
will observe confirmation of the Born rule, or that very probably
all observers will observe confirmation of the Born rule.  
On the contrary, many observers in an Everettian multiverse
will definitely observe convincing {\it disconfirmation} of the Born rule.  
Nor can one look at Everettian quantum theory and 
conclude that any given observer in the multiverse
will probably observe confirmation: the theory has  
no notion of standard probability available to even make sense of any
such claim.   And if the theory doesn't explain the data, the data
don't support the theory.  

There seems to be no good way around this, and if so, then 
that's the end of Everettian quantum theory as a serious contender: 
a theory with no predictive power {\it should} lose the scientific competition
against theories that predict what we actually see.   
However, Greaves and Myrvold \cite{greavesmyrvoldvol} have offered an 
attempt at a solution, by giving a general account purporting to
explain why agents who take seriously the possibility of many-worlds
theories can use observational data to confirm particular theories 
and refute others.  
Their account is illuminating, and raises
some very interesting questions about many-worlds theories.
Ultimately, though, it seems to me that it does not show, as
claimed, the possibility of explaining our observations from
a many-worlds theory and thus confirming one many-worlds theory
against another.  Rather, it highlights some apparently insuperable    
problems that prevent us from doing so.
As Greaves and Myrvold's arguments are set 
out in detail elsewhere in this 
volume, in this discussion I will simply summarize the implications 
of their confirmation algorithm in toy models, and point out the 
problems that arise.  

\subsection{The problem of inappropriate self-importance} 

It suffices to consider very simple many-worlds theories, 
containing classical branching worlds in which the branches
correspond to binary outcomes of definite experiments.  
Consider thus the {\it weightless multiverse}, a 
many-worlds theory of type $CBU_1$, in which the
machine produces only two possible outcomes, writing $0$ or $1$
onto the tape.   Recall that in $CBU_1$ there is no fact of the
matter about weights attached to the branches containing $0$ outcomes
and $1$ outcomes, although the inhabitants think there may be. 
This is the many-worlds analogue of an indeterministic one-world 
theory containing a sequence of binary experimental outcomes which 
are not only not determined but also not governed by any probabilistic
law.   Suppose now that the inhabitants begin a series of experiments
in which they push the red button on the machine a large number, $N$,
times, at regular intervals.   Suppose too that the inhabitants
believe (correctly) that this is a series of independent 
identical experiments, and moreover -- this is not essential, but
simplifies the discussion -- believe this {\it dogmatically}: no 
pattern in the data will shake their faith.  Suppose also that they 
believe (incorrectly) that their multiverse is governed by a many-worlds
theory with unknown weights attached to $0$ and $1$ outcomes, identical
in each trial, and seek to discover the (actually nonexistent) values
of these weights by following Greaves-Myrvold's learning algorithm.

After $N$ trials, the multiverse contains $2^N$ branches, corresponding
to all $N$ possible binary string outcomes.   The inhabitants on a 
string with $pN$ zero and $(1-p)N$ one outcomes will, with a degree
of confidence that tends towards one as $N$ gets large, tend to
conclude that the weight $p$ is attached to zero outcome branches
and weight $(1-p)$ is attached to one outcome branches.   In other
words, everyone, no matter what outcome string they see, tends towards
complete confidence in the belief that the relative frequencies
they observe represent the weights.   

Let's consider further the perspective of inhabitants on a branch with $pN$ zero
outcomes and $(1-p)N$ one outcomes.   They do not have the delusion
that all observed strings have the same relative frequency 
as theirs: 
they understand that, given the
hypothesis that they live in a multiverse, {\it every} binary string,
and hence every relative frequency, will have been observed by someone.
So how do they conclude that the theory that the weights are $(p, 1-p)$
has nonetheless been confirmed?   Because, following Greaves-Myrvold's
reasoning, they have concluded that the weights measure the {\it importance}
of the branches for theory confirmation.   Since they believe they have
learned that the weights are $(p,1-p)$, they conclude that a branch
with $r$ zeroes and $(N-r)$ ones has importance $p^r (1-p)^{N-r}$.  
Summing over all the branches with $pN$ zeroes and $(1-p)N$ ones, or 
very close to those frequencies, thus gives a set of 
total importance very close to $1$; the remaining branches have 
total importance very close to $0$.   So, on a set
of branches that dominates the importance measure, the theory that
the weights are (very close to) $(p, 1-p)$ is indeed correct.
All is well!  By definition, the important branches are the ones that
matter for theory confirmation.   The theory is indeed confirmed!  

The problem, of course, is that this reasoning applies equally well
for all the inhabitants, whatever relative frequency $p$ they see
on their branch.   All of them conclude that their relative frequencies
represent (to very good approximation) the branching weights.  
All of them conclude that their own branches, together with those
with identical or similar relative frequencies, are the important ones
for theory confirmation.   All of them thus happily conclude that their
theories have been confirmed.   And, recall, all of them are wrong:
there are actually no branching weights.  

\subsubsection{Comparison with the one-world case}

It's illuminating to compare the case of an inhabitant of the analogous
one-world universe, in which pressing the red button produces either a
$0$ or a $1$ on the tape but there is no law, either deterministic or 
probabilistic, governing these outcomes.   After $N$ experiments in 
which he sees $pN$ zeroes and $(1-p)N$ ones, he tends towards confidence
in the theory that zeroes have probability $p$ and ones have probability
$(1-p)$.   

Let us again restrict attention to theories --- in this case probabilistic
one-world theories --- that dogmatically assume the 
experiments are identical and independent. 
Among such theories, the selected theory does indeed characterize, better
than all its competitors, all the relevant data in the universe ---
i.e., all the outcomes of the $N$ experiments.     
Of course, further data could change that conclusion.   
But, so long as we consider only the 
relevant data, it's something of a puzzle to pin down whether
it's wrong to adopt the theory {\it pro tem}, and if so
precisely why.  Is there a physically
meaningful sense in which a universe that 
{\it looks} as though it contains data resulting from
a sequence of independent identical coin tosses with a
probability $p$ of outcome zero is
distinct from one that {\it does} contain such data?  And
if so, how precisely should we characterize the distinction?  

On the view of physical randomness 
discussed in section \ref{oneworldprob}, the answer
to the first question is no.  In any case, 
however one answers the questions, it seems that  
any possible error here must be subtler than and distinct from
the error highlighted above in the many-worlds case.
In the many-worlds case, recall, all observers are aware
that other observers in worlds with other data must exist, but each is led to 
construct a spurious measure of importance that favours 
their own observations against the others', and this leads to an obvious 
absurdity.   In the one-world case, 
observers treat what actually happened as important, and ignore 
what didn't happen: this doesn't lead to the same difficulty.  

\subsubsection{Numbers in the sky}

Consider next the {\it decorative weight multiverse}, a type $CBU_2$
variant of the weightless universe. 
This universe has a constant of nature fixed by the technologically advanced 
beings, a real number $p$, with $0 < p < 1$. 
As before, whenever the red button is pressed in 
a simulated universe, that universe
is deleted, and successor universes with outcomes $0$ and $1$
written on the tape are initiated.   
This time, the technologically advanced beings 
also write the numbers $p$ and $(1-p)$ in an 
inaccessible part of the skies of the $0$ and $1$ 
successor universes,  respectively. 
These numbers are visible to the inhabitants, but have no other
physical significance.   

There is thus a formal sense in which distinct weights are attached to 
the $0$ and $1$ branches.   However, by hypothesis, these weights
are decorative: there are no rational grounds for assigning them
any fundamental physical meaning or any r\^{o}le in constraining
rational actions.   We can thus run through a discussion of theory
confirmation precisely parallel to that for the weightless multiverse.  

This illustrates again that the mere fact that
Born weights are mathematically defined in Everettian quantum theory 
does not {\it per se} justify assigning them any role in theory
confirmation.  They could be merely decorative.    
  
\subsection{Separating caring weights from theory confirmation}

To investigate further it's helpful to consider 
branching world models in which
there {\it are} weights attached to the branches, in such a way
that the weights could plausibly be regarded as important for
making rational decisions.   I want here to tell specific stories
about the weights, in order to illustrate a crucial distinction
between two possible definitions of importance. 

\subsubsection{The replicating multiverse}

Consider first the {\it replicating multiverse}, 
a multiverse of type $CBU_3$ with a machine like the one above, in 
which the branches
arise as the result of technologically advanced beings running simulations.   
Whenever
the red button is pressed in a simulated universe, that universe
is deleted, and successor universes with outcomes $0$ and $1$
written on the tape are initiated.   Suppose, in this case, that
each time, the beings create {\it three} identical simulations
with outcome $0$, and just one with outcome $1$.  From the 
perspective of the inhabitants, there is no way to detect that 
outcomes $0$ and $1$ are being treated differently, and so 
they represent them in their theories with one branch each.   In fact, though, 
given this representation, there is an at least arguably natural 
sense in which they ought to assign to the outcome $0$ branch three
times the importance of the outcome $1$ branch: in other words, 
they ought to assign branch weights $(\frac{3}{4}, \frac{1}{4} )$.  

They don't know this.  But suppose, as before, that they believe
that there are unknown weights attached to the branches, and 
follow the Greaves-Myrvold procedure for identifying those weights.
What happens now?   After $N$ runs of the experiment, there will
actually be $4^N$ simulations -- although in the inhabitants'
theoretical representation, these are represented by $2^N$ branches.
Of the $4^N$ simulations, almost all (for large $N$) will contain
close to $\frac{3N}{4}$ zeroes and $\frac{N}{4}$ ones.   
These simulations will contain inhabitants who, following 
Greaves-Myrvold, believe they have confirmed that the branch 
weights (in their own theoretical representation, which remember
contains only $2^N$ branches) are very close to $(\frac{3}{4}, \frac{1}{4} )$.
They believe too that the weights define an importance measure on
the branches: a branch with $r$ zeroes and $(N-r)$ ones has 
importance (very close to) $ ( \frac{3}{4} )^r ( \frac{1}{4} )^{N-r}$.   
They thus conclude that their weight assignment will be confirmed on 
a set of branches whose total importance is close to $1$.  

Now, I think I can see how to run some, though not all, of an argument
that supports this conclusion.    The branch importance measure defined 
by inhabitants who find relative frequency $\frac{3}{4}$ of zeroes 
corresponds to the counting measure on simulations.     
If we could argue, for instance by appealing to symmetry, that 
each of the $4^N$ simulations is equally important, then this
branch importance measure would indeed be justified.   
If we could also argue, perhaps using some form of anthropic
reasoning, that there is an equal chance of finding oneself in any of 
the $4^N$ simulations, then the chance of finding oneself in a
simulation in which one concludes that the 
branch weights are (very close to) $(\frac{3}{4} , \frac{1}{4} )$ 
would be very close to one.  Turning that around, the theory that the
branch weights are $(\frac{3}{4} , \frac{1}{4} )$ would then imply that,
with high probability, one should expect to see relative frequency
of zeroes close to $\frac{3}{4}$.   
There would indeed then to be a 
sense in which the branch weights define which subsets of the branches
are important for theory confirmation.   

It seems hard to make this argument rigorous.  In particular, the 
notion of ``chance of finding oneself'' in a particular simulation 
doesn't seem easy to define properly.   Still, we have an arguably natural
measure on simulations, the counting measure, according to which
most of the inhabitants will arrive at (close to) the right theory
of branch weights.   That might perhaps be progress.   

\subsubsection{The qualia enhancing multiverse}

But consider now the {\it qualia enhancing multiverse}, 
again a multiverse with the same 
type of machine, in which the branches
arise in the way we've previously considered, as the result of 
technologically advanced beings running simulations.   Whenever
the red button is pressed in a simulated universe, that universe
is deleted, and successor universes with outcomes $0$ and $1$
written on the tape are initiated.   This time, though, 
the beings create just one simulation with outcome $0$, and one with 
outcome $1$, but devise their simulations so that the qualia --
the mental sensations -- of the inhabitants in the outcome $0$
simulation are three times as intense.   
As before, from the perspective of the inhabitants, there 
is no way to detect that outcomes $0$ and $1$ are being treated 
differently, and so 
they represent them in their theories with one branch each.   

There is, again, an arguably natural 
sense in which they ought -- if they were aware of the rules of
their multiverse --
to assign to the outcome $0$ branch three
times the importance of the outcome $1$ branch: in other words, 
they ought to assign branch weights $(\frac{3}{4}, \frac{1}{4} )$.  
Recall, pleasure and pain in outcome $0$ branches have tripled in 
intensity.   The welfare of successors on outcome $0$ branches is
felt more intensely, and in that sense it matters more.

Let me deal here with three possible objections: 

\begin{itemize}

\item[(a)]  It might be argued that qualia enhancement should be
analysed differently, as an example of an unannounced alteration in 
utility functions: the actual payoff
of winning a bet with outcome $0$ is three times the expected payoff,
since the inhabitants don't expect any qualia enhancement.  
Certainly it {\it could} be analysed in this way.  But this 
reflects an arbitrary choice that always needs to be made in many-worlds
theories.   (Precisely the same argument
could be made in the case of the replicating multiverse, for example.) 
The statement that one branch is $N$ times as important as another can 
always be recast as a statement that utilities on the first branch
are rescaled by $N$ relative to those on the second.   
So, we can legitimately analyse qualia enhancement as an effect altering 
the relative importance of branches, and it's interesting to do so, 
as this lets us test 
general propositions about the confirmation of theories attaching
importance to branches.  

\item[(b)] The reader may
not believe that there is a sensible account of experience
involving qualia, or that intensifying qualia makes any sense.
Never mind.  It's just a useful device to make a point about
branch measures.   It could be formulated in another way: 
we could suppose that the simulators arrange that all bets 
have payoffs with three times the expected utility on
outcome $0$, while erasing the relevant bits of the inhabitants'
memories so that they're not aware that the payoff tripled.  

\item[(c)] One might also worry that 
inhabitants in an outcome $0$ branch would notice that the  
intensity of their qualia has just tripled.   For the sake of
the argument, we must assume not.   Insofar as the notion
of qualia enhancement makes sense, this seems reasonable: their memories
will triple in intensity along with everything else.
\end{itemize}

Suppose, once again, that the inhabitants believe
that there are unknown weights attached to the branches, and 
follow the Greaves-Myrvold procedure for identifying those weights.
What happens now?   After $N$ runs of the experiment, there will
be $2^N$ simulations -- now correctly represented by 
$2^N$ branches in the inhabitants' many-worlds theory.
The simulations will contain inhabitants who, following 
Greaves-Myrvold, believe they have confirmed that the branch 
weights are very close to $(p, 1-p)$, because their observed
relative frequency is $p = r/N$, for each $r$ in the range
$0 \leq r \leq N$.  They believe 
that the weights define an importance measure on
the branches: a branch with $r$ zeroes and $(N-r)$ ones has 
importance (very close to) $ ( p )^r  (1-p )^{N-r}$.   
They thus conclude that their weight assignment will be confirmed on 
a set of branches whose total importance is close to $1$.  
Now, in one sense, the inhabitants whose observed relative frequency
$p = 3/4$ are a special case.   Their inferred importance measure
equals the natural importance measure defined by qualia intensity.  
And if we weight the branches by this importance measure,
it is the case, by the same calculation as before, that, on 
a set of branches with total measure close to one, the inhabitants 
end up with (very close to) the "right" branch weights, 
$ ( \frac{3}{4}, \frac{1}{4} )$. 

But wait!   If we count the simulations, the inhabitants who
arrive at weights $ ( \frac{3}{4} , \frac{1}{4} )$ are a tiny
minority.   Almost everyone arrives at the wrong branch weights --
and, as in our earlier example, almost everyone arrives at a measure
of importance according to which branches with (very close to) their
observed relative frequency are the important ones.  By the natural
simulation counting measure, theory confirmation has spectacularly
failed.  
 
What these last two examples show is that there are two distinct senses,
which Greaves-Myrvold and Wallace fail to separate, in which a 
branch weight might
possibly be said to be a measure of importance.   It could be said 
to be a "caring measure", if there is some reason to care 
differently about the welfare of successors on different branches.
And it could be said to be, for want of a better term, 
an "explanatory counting measure", if there is some reason
to think that we are likelier to {\it find ourselves} on some branches
rather than others -- or some other argument to show 
that a branching theory which predicts the observed relative
frequencies (or other data) on a set of branches of high 
explanatory counting measure thereby explains them. 
What we've seen is that the first property doesn't necessarily imply
the second, and it's the second that is 
needed for an adequate account of branching theory confirmation.

Couldn't a many-worlds theorist then simply {\it postulate}
the existence of an explanatory counting measure?   (And perhaps also  
postulate that a caring measure exists and equals the explanatory
counting measure?) 

A preliminary remark: even postulating a caring measure -- which 
{\it has} been 
proposed\cite{papineauvol} in the Everettian literature -- already
seems a very strange manoeuvre.  Physical theories can certainly give 
reasons for rational agents to perform certain actions if they
have certain goals.   But what's envisaged here is a theory
that {\it by fiat} imposes a constraint on rational behaviour.  I'm 
not clear -- and at least some Everettians (e.g. \cite{saundersvol})
seem to share this worry -- 
that this makes any sense, either as an idea about
physics, or about rationality.\footnote{See Appendix \ref{rationalconstraint}
for further discussion.}

In any case, when it comes to postulating an explanatory
counting measure, one should be clear: the proposal is  
that a many-worlds theory {\it defines}, by fiat, without any attempt
at further justification, whose observations matter and whose may
be neglected, when it comes to testing and confirming the theory.  
The theory defines its own -- highly 
non-standard -- criteria for deciding whether or not it is
a scientific success.  

One could play this sort of game, of course, even in one world.  
For example,
Alice could define a theory that includes -- as a postulate, with no further 
explanation -- the principle that everyone who agrees 
with her observations and her theoretical interpretation is important
for theory confirmation, and everyone else is negligible.   
She could then announce, after checking with the important people, that 
her theory is confirmed.
This would be self-consistent, and maybe politically adept, but it 
wouldn't be science.  

It's no more scientifically respectable to declare that we can, without
further justification, confirm Everettian quantum theory by 
neglecting the observations made on selected low Born weight branches.
A Pavlovian association of low Born weight 
with small probability --  illegitimately carried
over from one world quantum theory -- may perhaps lend an
aura of greater respectability.   
But in Everettian quantum theory the Born weight is 
simply a number attached to branches.  
It has no intrinsic relevance to theory confirmation, and unless we add 
further structure to the theory, we cannot justify assigning it any such 
role.   

Note again the contrast here with the one-world case:
one-world probabilists do not pick and choose which observations 
are to be used for theory generation or confirmation.  

\subsection{Many-worlds confirmation: conclusion}

To explain how we could come to confirm Everettian many-worlds
quantum theory it is not enough to note that we have Born
weights to hand and so can automatically give them a 
confirmation-theoretic r\^{o}le.   As the decorative weight multiverse 
illustrates, branch weights can be simply irrelevant to 
theory formation and confirmation.   
  
Nor can Wallace's arguments for treating the 
Born weights as a caring measure suffice, even if we take Wallace's
result at face value.  As the qualia enhancing multiverse
illustrates, a caring measure is not necessarily an explanatory 
counting measure.  

Thus, the most sympathetic (though unauthorised) 
translation of Greaves-Myrvold's 
account of many-worlds and confirmation that I can find requires us
to add structure that {\it justifies} the existence of an 
explanatory counting measure.   This requires interpreting 
Everettian quantum theory, along with competing many-worlds theories
theories, as modelled by versions of the replicating multiverse,
with branches constantly being deleted, and successor branches
created.  We need to postulate that the number of 
simulations or realisations of a given branch at a given 
time is proportional to the branch weight, and to assume 
that it is rational to treat all realisations as equally valuable.
We need also to postulate something like an
anthropic principle that tells us that, in some sense that
needs to be properly defined, the chance of finding ourselves in 
one of a given class of realisations at a given time is proportional 
to the number of realisations in the class.  

This, if it could be made rigorous, would suggest something resembling 
the objectively determinist 
``momentary minds'' version of Albert-Loewer's many-minds 
interpretation\cite{almanyminds,bellmw,barbour}, 
in which the minds exist only instantaneously, with no continuous
identity extending over time.   This isn't a picture I find easy to take
seriously.    As I read them, none of the Everettian 
contributors to the present volume would 
wish to defend this account -- and yet it seems very closely aligned with 
some of their intuitions.  Let me close here by inviting 
readers to see if they can find a better
way of rigorously justifying Greaves' gloss:\cite{greaves}

\begin{quotation} 
But since we have
a measure over our successors, we can, if we find it intuitive, talk of 'how
much successor' sees spin-up. I have a preference for my spin-down successor
to receive chocolate, rather than my spin-up successor, because there is more of
the former; more of my future lies that way. Thus, I think, Lockwood's (1996)
talk of a 'superpositional dimension', and/or Vaidman's (1998, 2001) suggestion
that we speak of the amplitude-squared measure as a 'measure of existence', are
somewhat appropriate (although we are not to regard lower-weight successors
as less real, for being real is an 
all-or-nothing affair -- we should say instead
that there is less of them).
\end{quotation}

\section{Fuzziness, rationality and decision theory in many worlds}

Two of the most interesting recent developments in the Everettian
literature, in my view, have been the attempt to argue for
an intrinsically fuzzy emergent quasiclassical ontology\cite{wallacevolone}
and (as already discussed) the attempt to reinterpret
Born weights via a many-worlds version of 
decision theory\cite{wallacevoltwo}.   
Interesting, but flawed -- 
each project has deep problems, and they appear to be based on
inconsistent premises.   

\subsection{Fuzziness and its limitations}

Granted, as Wallace\cite{wallacevolone} notes, viable higher level scientific 
theories can and do, indeed, supervene on more fundamental theories. 
Objects in those theories need not have any unique and
precise definition in terms of fundamental concepts: 
there is, indeed, no unique, natural, precise, chemical characterisation 
of a tiger. 

Nonetheless, there is a very strong reason
for seeking\cite{akcritique} a precise mathematical
formulation of the intuition that many branching worlds emerge from 
unitary quantum theory -- or else a precise mathematical formulation
of some other structure consistent with Everettian ideas -- 
namely, that it is not at all clear that,
without such a formulation, we have a well-defined scientific theory
to discuss.   (This, it seems to me, is why both  
Everettians\cite{graham, deutschone} and
critics\cite{bellmw,almanyminds} have often attempted to find  
mathematical structures that might explain the notion of 
branching.)    The alternative 
strategy, proposed by Wallace\cite{wallacevolone}, of trying 
to interpret the implications of 
a fundamentally mathematical theory in terms of 
higher level fuzzily defined constructs carries a very obvious
danger --- namely, a retreat into vagueness and hand-waving on points where 
precision really is required.  
It's hard to run a serious argument (pro or con), let alone prove a 
rigorous theorem, if one doesn't, in the end, know quite what 
one's talking about.   

\subsection{Fuzzy minds}

A case in point is Wallace's appeal to functionalist intuitions in 
trying to give an account of the 
mind states of agents in Everettian quantum theory.
Readers are, I think, owed a much more precise explanation of what, actually, 
is supposed to follow from this, since
some rather crucial points appear to turn on unspecified details. 

For instance, on this account, do 
distinct mind states necessarily correspond to orthogonal quantum states?
If so, wouldn't this account necessarily supply us with a 
preferred orthogonal decomposition of the unitarily evolving 
quantum state?   And wouldn't this, pace Wallace\cite{wallacevolone}, 
allow a precise definition of a relevant branching structure after all?   

Wallace places great emphasis on the lack of a 
unique natural definition of a quasiclassical branch, and hence
the impossibility of agents formulating a rational strategy based
on counting distinct future branches.    
But it's at least as relevant to examine whether our account 
of mind states supplies a natural definition of a future self, and 
whether it might be possible for agents to formulate a rational 
strategy based on counting distinct future
selves?   Can't an agent identify successor selves as distinct if and 
only if they have distinct mind states, ascribe to distinct successors
a branching history corresponding to that recorded in their memories,
and use {\it those} data to define a rational strategy for taking account
of their welfare? (These points are pursued further in appendix
\ref{countdesc}.)  

On the other hand, if non-orthogonal quantum states {\it could} correspond to 
distinct mind states, how would we even begin to 
connect quantum theory with even the appearance of probabilities?   
Quantum theory gives no general rule to calculate a 
probability of a transition from an 
unknown state belonging to one fuzzily defined set of states 
(corresponding to mind state A) to an unknown state belonging to another
(corresponding to mind state B).   But that's what we'd need to calculate,
in principle, in order to obtain a number corresponding to the apparent 
probability of arriving at state B when starting in state A.   
Maybe one could cook up such a rule, and then explain how the Born
rule emerges as an approximation under suitable circumstances -- but
it's not obvious how, and this would certainly be going beyond 
quantum theory as presently understood.  

Both options thus lead to serious, perhaps  
insuperable, difficulties. 

\subsection{Can precise preferences arise in a fuzzy ontology?}

Another very basic worry about Wallace's programme is its 
equivocation over mathematical rigour. 
Everything in Wallace's ontology that's 
relevant to rational decisions --- including agents, 
the quasiclassical branches they inhabit, the branch states, 
and the branch Born weights, and the distinction between micro-states
and macro-states --- is intrinsically fuzzily defined\cite{wallacevolone}.   
There is,
on Wallace's account, no precise fact of the matter about the 
different quasiclassical states that would result after a bet
on a quantum experiment, nor about the Born weights of the branches
corresponding to those quasiclassical states.   And this isn't merely
because quantum theory doesn't supply a unique natural definition of 
elementary branches and branching events: the {\it total} Born 
weight of all the quasiclassical branches describing a spin-up
outcome of a Stern-Gerlach experiment isn't precisely defined either. 

Now, to be sure, the total weight {\it is} supposed to 
be approximately defined.  We are supposed, on Wallace's
account, to be able to say that it's in a range of the 
form $ R = ( p - \epsilon, p + \epsilon )$, where $\epsilon$ is 
very small, and $p$ thus 
represents an approximate total Born weight.\footnote{That
all ambiguities in total weights of quasiclassical outcomes 
are necessarily very small seems plausible and is what Wallace expects.  
Given the level of conceptual imprecision in discussing the emergence
of quasiclassical structures from unitary quantum theory, 
though, it is hard to be certain even of this.}
But we're not supposed to be able, on
this account, to reduce $\epsilon$ to zero: below some level 
of precision, it becomes unavoidably arbitrary, just a matter of taste 
in your choice of branch definition, whether you take  
the total weight as $p_1 \in R$ or $p_2 \in R$. 

And yet, Wallace's decision theoretic programme postulates
that each rational agent should have a {\it precisely specified} 
and {\it complete}
preference ordering among a very large class of
possible unitary maps that produce different possible 
future global states.  Where could such
a preference ordering possibly come from?  
The ordering is supposed to be agent-dependent.  Physics doesn't equip rational
agents with some personal preference ordering on global states: they
have to arrive at their preferences by introspection and reasoning.
If one accepts Wallace's conclusions, the only ultimately relevant
quantities are branch weights and the agent's personal utilities for 
macrostates (whose existence is supposed to follow given the preference
ordering axioms).
But even a super-agent who finds they can calculate the former and can
identify the latter by pure introspection would find these quantities 
only fuzzily defined -- so that, in comparing
some pairs $( U_1 , U_2 )$ of actions on a given state $\ket{ \psi }$, 
however hard they try and however carefully they analyse the alternatives, 
they wouldn't be able to identify a reliable preference, {\it not}
because the resulting global states are precisely equivalent, but 
because their difference is fuzzily ambiguous.
On some views, $U_1 \ket{ \psi }$ would seem very slightly preferable;
on others, $U_2 \ket{ \psi }$ would.  
In Wallace's notation \cite{wallacevolone} for preference orderings, 
neither $U_1 \succeq_{\psi} U_2$ nor $U_2 \succeq_{\psi} U_1$ would hold in all
ways of looking at the situation.  
Nor does it seem legitimate to postulate that $ U_1 \sim_{\psi} U_2 $ 
must hold in such cases.  One can imagine the possibility
of a sequence $( U_1 , \ldots , U_n )$ such that no 
preference can reliably be identified
between $U_i \ket {\psi }$ and $U_{i+1} \ket{\psi}$, for $i=1 , \ldots, (n-1)$,
but nonetheless setting $U_i \sim_{\psi} U_{i+1}$ violates transitivity,
since $U_1  \succ_{\psi} U_n $ {\it does} hold no matter what 
view the agent adopts of the fuzzy facts.   

We are not, in any case, super-agents, and
can only read Wallace's arguments as prescriptions
for ideal rationality rather than descriptions of our real-world behaviour.
None of us in fact has a complete and precise preference 
ordering among the relevant unitaries.   Wallace, in effect,
is telling us that we should ideally adjust our
reasoning and their behaviour so as to be consistent with 
some complete preference ordering.   But how?  There is no 
natural algorithm available: any choice will involve uncountably
many arbitrary decisions on pairs of preferences.\cite{savageworry}   
And why?  Given that no 
choice of ordering will have any intelligible justification, 
even after the entire analysis is complete, how can there 
be a rational compulsion to make some choice (even if, 
counterfactually, it were practical)?

Here, it seems to me Wallace's prescription runs into essentially the
same difficulties that he identifies in other ways of thinking about 
Everettian branching.  One {\it could}, in principle, find some 
(perhaps ad hoc) prescription defining a branching structure for
the unitarily evolving state vector, and one could then use this
structure to define a rational Born-rule-independent strategy based 
on branch counting.  Wallace accepts that such a strategy is
not logically inconsistent, but argues that it is 
likely to be difficult to implement in practice (because defining
a precise branching structure is difficult) and hard to justify
in principle (because the definition seems to require ad hoc choices).  
Both objections apply -- arguably with at least equal force -- to
the Wallace programme.  

This also reinforces the point that the case for Wallacean rationality cannot 
possibly rely on the lack of any practical alternative strategy. 
A very practical alternative is to follow whatever 
combination of instinct and reasoning evolution provided us before
we became aware of Everettian quantum theory.  Altering that strategy so as to
comply rigorously with Wallace's axioms isn't practical; even coming close
to doing so may not be.   To be persuaded that we ought
to try, we would need to be rationally persuaded not only that we should ideally
be Wallaceans, but also that there is a practical method which allows
us to become closer to being Wallaceans, and that we will be better off
if we employ this method.\footnote{And to run such an argument, Wallaceans 
would, inter alia, need to find suitable precise definitions 
of ``closer'' and ``better off''.}

In short, 
given Wallace's account of a fuzzy ontology, there seems a 
strong reason to doubt Wallace's most basic
postulate of rationality, $R1$, which states that rational agents have
a {\it complete} (or {\it connected}) preference ordering on the 
unitary operations available to them at any state $\ket{\psi}$.   
No actual agent in a fuzzy Everettian ontology will ever be able to 
arrive at such an ordering in practice.  Moreover, even if they had infinite 
computational power, fixing an ordering would require making a very
complicated ad hoc choice which can have no complete rational justification.  
Yet without $R1$, the purported derivation of the Born rule\cite{wallacevoltwo} 
fails at the first step. 

It's not clear to me that there is any fix for this, but let me
comment briefly on two possible responses. 

First, one might perhaps try weakening the postulate $R1$ to suggest that agents
have, or should aspire to have, a preference ordering that 
{\it approximates} a complete ordering, in the hope of then proving that
their policy should approximate Born-weighted mean
utilitarianism.  One problem with this is that one would need first
to find and justify a suitable definition of approximation applied to
preferences between pairs of unitary operations.   As these
are unquantified binary relations, it doesn't seem obvious that any
suitable definition exists. 

Second, one might consider the desperate resort of {\it postulating} 
a total ordering as part of the physical theory.  But even that surely 
isn't available here.  The orderings, recall, are
agent-dependent, and even the most postulate-happy Everettian would surely
recoil from requiring that {\it fundamental physical laws} specify 
independently, agent by agent, the preferences of every agent 
instantiated in nature.   

Trying to formulate a rigorous decision theory for preferences in a 
fuzzy ontology may thus be rather like trying to build a skyscraper on 
mud.  

\subsection{Circularity of the Wallace programme?} 

Zurek\cite{zurekvol} flags another worry about the
logical relation between the two parts\cite{wallacevolone,wallacevoltwo}
of Wallace's programme, namely an apparent circularity.   
Wallace envisages a fuzzy quasiclassical ontology
arising as the result of mathematical regularities observable within
components of the unitarily evolving universal wave function.  
These regularities are supposed, in a realistic cosmological model,
to arise through the decoherence of classical variables and to be
defined by what Gell-Mann and Hartle term a quasiclassical 
domain \cite{hartlevol}, in which, for example, operators
approximately quantifying local mass densities approximately 
follow classical equations of motion with probability close to one.   
Here the probability for a history defined by a sequence of 
operators $P_1 (t_1) , \ldots , P_n (t_n )$ is given by the 
decoherence functional
\begin{equation}
\Tr ( P_n (t_n ) \ldots P_1 (t_1 ) \rho_{{\rm initial}} P_1 (t_1 ) \ldots
P_n (t_n ) ) \, . 
\end{equation}

In other words, the ontology is {\it defined} by applying the Born rule.   
Even if one could show, as Wallace claims, that agents defined within that 
ontology are rationally justified in using the Born rule as a 
calculus for decisions, it would seem incorrect to portray this
argument as a {\it derivation} of the Born rule within Everettian
quantum theory.   Wallace's argument should rather be understood
as attempting to show something weaker: that the Born rule re-emerges as output
(albeit, to be fair, in an interesting and non-obvious way) 
if assumed as input.  
Even if correct, this would leave 
open the possibility that there are many different 
consistent and essentially inequivalent ways of defining ontologies 
that include distinct types of agents for whom different rational 
decision calculi can be established.
It would thus fail to explain whether and (if so) why our own decision calculus
should be based on the Born rule.  It
would also leave open the questions as to whether and (if so) how   
agents in some consistently defined Everettian ontology can arrive 
at the rational decision calculus appropriate to their ontology. 

\subsection{Problems with Born-weighted mean utilitarianism} 

Wallace\cite{wallacevoltwo}, developing earlier 
ideas of Deutsch\cite{deutschtwo}, partly 
in response to criticisms (e.g. Ref. \cite{barnumetal}) of 
the latter, 
then goes on to argue that from a few simple and purportedly 
natural axioms we can prove that  
rational agents who believe 
themselves to be in a universe described by many-worlds quantum theory
are rationally required to (a) have a utility function that quantifies 
the value they assign to possible future quasiclassical events,
(b) act so as to maximise their Born-weighted mean utility.

As we just saw, Wallace's first postulate, $R1$, seems to run
into a fundamental obstacle, since neither Born weights nor 
quasiclassical histories (and thus their utility) are precisely
defined in his ontology, and without $R1$ the decision theoretic 
argument, which, inter alia, implies the existence of a utility function, 
fails.   Moreover, even for an agent who {\it has}
a utility function applicable to all relevant quasiclassical
histories, the strategy of maximising Born-weighted mean utility
is not well-defined.  For a real world agent in state $\psi$
there will generally be available unitaries $U_1$ and $U_2$ for 
which it's a matter of arbitrary definitional choice 
whether $U_1$ or $U_2$ produces higher Born-weighted mean utility. 

There's a further practical problem, which isn't apparent in simple
models of many-worlds experiments but is a serious worry in 
realistic applications.  To be a rigorous Born-weight
mean utilitarian in the real world, one must allow for the possibility of   
small Born weight branches with extreme negative or positive utility.
The mean Born-weighted utility of a bet that, with 
Born weight close to $1$, involves 
small utility gains or 
losses, is radically altered if it also creates a $10^{-25}$ Born weight 
branch of utility $-10^{30}$.   Now, the Deutsch-Wallace-Savage arguments imply
no bounds on agents' utility functions.   It seems unlikely that
any {\it a priori} argument can supply one, since pure rationality 
imposes no bound on utility functions -- and in practice, for example,
there seems to be no generally agreed lower bound on the utility cost 
assigned to the destruction of the Earth or similar catastrophes\cite{akrisk}. 
A rigorous real world Born-weight mean utility calculation 
thus typically requires very careful analysis of small weight branches.   
In fact, even ensuring that the sum defining the mean 
utility {\it converges} requires
careful analysis of small weight branches: consider, for example,
the possibility of a set 
of branches of weight $2^{-n}$ and utility $-3^n$ for all integers $n \geq N$.  

Practically speaking, the best that real world agents are likely to be able
to do is first simplify their model, by excluding events below
some weight threshold, and then estimate a Born-weighted mean 
utility within that model --- with no assurance that the estimated 
mean utility is close to the true mean utility (if indeed the
latter exists).\footnote{
Given the fuzzy ontology, we should more accurately
say ``to any possible assignment of the imprecisely
defined value of the true mean utility''.   For the sake of 
readability, we take this qualification as read in what follows.} 
This needs emphasising, since much of Wallace's case 
against alternative rational strategies is based on the claim 
that they are ill-defined or impractical or both.  
Actually, as we will see, alternative
strategies can sometimes
be rather {\it better} defined and {\it more} practical 
than Born-weight mean utilitarianism. 

\subsection{Everettian many-worlds rationality reconsidered}

\subsubsection{General remarks on life in a multiverse}

It seems {\it prima facie} surprising to claim that   
mathematical analysis could show that 
Born-weight mean utilitarianism, or any other strategy,
is the unique rational way of optimizing the 
welfare of one's own, and other people's, many
future selves in a multiverse.  
After all, human parents are faced with the not entirely disanalogous 
question of how to take into account the welfare of their genetic 
descendants in (most of us assume) a single world, and it's
a notoriously complex problem.  
People generally care not only about their descendants'
present welfare, but also about their expected future welfare after
our death.  They can, and sometimes do, frame guiding rules of thumb
to arbitrate between competing claims on their resources -- for
instance, to divide their estate equally among their children, or to
divide it according to their need.  They take into account their
children's relationships with one another, with others, and with society.
They tend to care about immediate descendants more than distant ones, 
in a way that generally follows no well-defined formula.  
Evolutionarily developed instincts also impel a more general concern
for our genes and those of the species.  This concern probably cannot
be precisely codified, but we can often find principles with which they are
roughly aligned and which roughly characterise the behaviour
they motivate.  For instance, some species' instinctive behaviour
might be roughly modelled as aiming to maximise an individual's 
expected number of descendants after $10^2$ years. 
Some humanists' aims might be modelled as aiming to maximise the 
survival probability of the human race (and its genetic successors) 
over the next $10^9$ years.  

Some of these principles require impossible
calculations to implement precisely, but can nonetheless legitimately 
be regarded as rational aims.  If we adopt them, we 
commit ourselves to trying to
satisfy them as best we can. In general, they imply conflicting
courses of action.  No one, I think, would seriously claim that any
one of them is uniquely rationally preferable to all the others.  We
just make decisions as best we can, imperfectly 
guided by logic, sometimes perhaps trying our best to 
optimise quantities we know we cannot properly calculate.  And we 
always did: before we were capable of rational reflection, evolution 
equipped us to muddle through, sometimes following one rule of 
thumb, sometimes another.  That's life.  Why should we expect 
evolution or rationality to have equipped 
us any better when faced with the bewilderingly underdetermined
imperative to care about our and everyone else's 
quantum descendants in a hypothetical multiverse?  

\subsubsection{Alternative Born-weight-sensitive strategies}

Suppose, for the sake of the discussion,
that we can somehow ignore the fuzziness of the ontology.
Suppose that we have an agent faced with a finite number
of choices $j$, each of which will create quasiclassical branches (although
not a unique quasiclassical branching structure) with well-defined
utilities $U^j_i$, in such a way that the set $S^j_i$ of branches 
with the same utility $U^j_i$ has a well-defined total Born weight $p^j_i$, 
and that the sums $\mu^j = \sum_i p^j_i U^j_i$ are finite.\footnote{
Without these assumptions, the mean utilitarian's strategy isn't
defined, in which case Wallace's argument has failed.}

Consider again some of the strategies listed in Sec. \ref{strategies}. 
The $x$-percentile utilitarian, for $0<x<100$, always has a well-defined
strategy, as does the future self democrat.   
The future self elitist and Price-Rawlsian's strategies are defined
provided that $\max_{j} \sup_i (U^j_i )$ and $\max_j \inf_i (U^j_i )$, 
respectively, are defined.  These will always hold true if 
the indexing set $I \ni i$ is finite.  They need not hold true
if the branch utilities are unbounded above or below (possibilities 
which are not usually considered by Everettians, and which perhaps
might be excluded by assumption, but possibilities nonetheless). 

As a practical matter, unless low Born weight extreme utility branches
can be excluded, the future self elitist and Price-Rawlsian
may have difficulty optimising their strategies, even if an optimal
strategy exists, since calculating $\sup_i (U^j_i ) $ or $\inf_i (U^j_i )$
requires analysing low Born weight branches that realise,
or converge towards, the extreme utility values.   This is also be a 
problem -- which may be easier or harder, depending on the details -- 
for the mean utilitarian.   Generically, it should not be a significant 
problem for the $x$-percentile utilitarian (for most $x$, say $1 < x < 99$), 
assuming the 
utility function is generically well-behaved over the range, since
the utility at the $x$-th percentile is then relatively insensitive 
to small perturbations of $x$, and so the calculation is
relatively insensitive to the details of
low Born weight extreme utility branches.  
It should also generally not be a problem 
for a purely self-concerned 
future self democrat, who would generally hope to be able 
to attain a majority decision without counting the votes from low Born 
weight extreme utility branches.\footnote{Of course, in both these last 
two cases, it {\it could} still be a problem if the numbers so conspire.}

\subsubsection{Some other strategies}

The {\it Gell-Mann--Hartle aesthete} fixes a particularly pretty 
quasiclassical consistent set $S$, which she uses to define a way 
of counting branches containing
her future selves.\footnote{I thank Hans Westman and Ward Struyve for
suggesting this example.}  Her quantum ontology is Everettian: she
agrees that her selected set has no fundamental physical significance.  
However, she thinks one needs {\it some} way of 
weighting future selves and that this one is as rationally
defensible as Born-rule-weighting or any other, and more aesthetically 
pleasing. 

The {\it value teleologist} fixes
a particular cosmological final density matrix $\rho_f$, whose
spectrum does not include zero.
In considering whether or not to accept a generalized bet on
a quantum experiment, or indeed making any decision dependent on a
quantum event, he uses pre- and post-selection, with some standard
theory of the initial cosmological conditions defining the initial
state $\rho_i$, and with $\rho_f$ defning the final state, in order to 
calculate the probabilities of the future worlds corresponding to the
possible outcomes.\cite{aharonovetal, gmhtwotime}  
He bets as if these were the actual probabilities.  This is not because 
he believes they are -- he believes in deterministic
unitary quantum mechanics and so doesn't think probabilities are
fundamental, and in any case
his physical theory is a standard cosmological theory with 
initial state $\rho_i$ and no post-selection on $\rho_f$.
However, for aesthetic or existential reasons, his interest in 
future events is conditional on the chosen final state post-selection.  

\subsubsection{Wallace's rationality postulates} 

We noted already that branch weight, branch macrostate, branch 
microstate, and reward are all only fuzzily defined in 
Wallace's ontology\cite{wallacevolone}, and that 
this gives strong reason to doubt Wallace's ordering axiom $R1$.   
It casts doubt too on whether the availability axioms $A3-A5$ and the 
rationality axioms $R3$, $R5$ even have a precise definition. 

Wallace's diachronic consistency axiom, $R2$, is violated
by the $x$-percentile utilitarian strategy, among others. 
Now, to be fair, one can find examples where the two conflict which
illustrate some motivation for diachronic consistency.   Consider the
possibility of being offered $N$ dollars
per unit time to stand in a radiation field, with a risk $p$ of lethality 
per unit time.  
An $x$-percentile utilitarian who considers this offer will
generally find their response depends on the timescale
over which they regard their decisions as binding: it could seem
a good offer considered as valid for the next second, and then good  
again for each successive second, but a bad offer if they have to
make a single decision about whether to accept for the next hour. 

Yet, even in this rather unusual example, the 
motivation for $x$-percentile utilitarians
is still clear when $x$ is close to $0$ or $100$, and
their actual strategy is intended as a practical approximation to 
their ideal strategy of Price-Rawlsianism or future self elitism. 
The Price-Rawlsian will decline unless the total risk is 
zero; the future self elitist will accept unless
the survival probability is zero.   
Note too that even here $x$-percentile utilitarianism 
{\it is} a well-defined strategy once a timescale for decisions is fixed. 
In the more normal circumstance of separated discrete decisions,
$x$-percentile utilitarianism seems both rationally defensible and 
practical, which suggests that the diachronic
consistency axiom is less rationally compelling than Wallace argues. 

Another reason to doubt $R2$, it seems to me, 
is that, pace Wallace's comment\cite{wallacevoltwo} -- 
\begin{quotation}
``In the presence of widespread, 
generic violation of diachronic consistency, agency in the Everett 
universe is not possible at all.'' 
\end{quotation} 
-- diachronic consistency actually {\it is}, strictly speaking, 
generically violated in real world decisions.  
A Savagean or Wallacean rational agent, recall, has to 
be equipped with a utility function as well as a probability measure
for outcomes.   Rationality is silent on the precise form of the utility 
function.  If we have one, it 
reflects our current values.
These generally change over time, as we do, 
partly as a result of decisions we have 
previously taken, whose outcomes affect
us in ways we cannot reliably predict beforehand: 
our own natures are too complex and too opaque to us, 
and we also change in response to our environment, which
is also complex and unpredictable.   The best 
it seems to me that one might hope to say of
diachronic consistency in real world decisions is that pretty often, in the 
short term, it approximately holds --- which clearly 
isn't a strong enough assumption to 
prove an interesting decision theoretic representation theorem.

Elga's proposal that Everettians might have a rational preference for 
future self diversity\cite{wallacevoltwoelga} also seems pertinent
here, as does the case for rationally preferring future 
society diversity.  (Why {\it not} 
exploit the scope for political compromise by causing   
society to evolve in different ways along different branches?) 
In both cases, it seems to me, 
contra Wallace\cite{wallacevoltwoelga}, diachronic 
consistency can be rationally violated.  
I can consistently believe now that it's a good thing that the 
global state should include future copies of me as a king and a beggar, 
while knowing that, if I ever find myself a beggar, I would strive to become
a king if I could.\footnote{Or perhaps, considering the uneasiness of 
crown-wearing heads, vice versa.}  From the perspective
of my future beggar self, the unpleasantness
of finding that {\it he} is the beggar outweighs the satisfaction of knowing
that diversity was achieved.   From my present perspective, the 
prospect of diversity nonetheless remains appealing. 

As Wallace himself notes earlier in his discussion\cite{wallacetwodictates}: 
\begin{quotation}
$\ldots$ to make a copy of myself and send him off to do a dangerous or 
disagreeable task -- and $\ldots$ to take actions designed to
prevent him shirking that task $\ldots$ is not {\it irrational}.
\end{quotation}
Indeed -- and this remains true if the task, for which I have a strong present
desire, is to ensure future self diversity.  The fact that my future 
selves will never interact makes no difference to the rational justification. 

Turning briefly to other postulates:
\begin{itemize}
\item[(R3)] Microstate indifference, $R3$, can be violated by value
teleologist strategies, among others.   
\item[(R4)] Continuity, $R4$, is 
violated by $x$-percentile utilitarian strategies,
among others.\footnote{Wallace\cite{wallacevoltwo} notes and discusses 
the special case of the Price-Rawlsian.}
\item[(R5)] Branching indifference, $R5$,
is violated by Gell-Mann--Hartle aesthete strategies, among others.  
\end{itemize}

\subsubsection{Summary}

Wallace argues that strategies other than
mean utilitarianism turn out, on closer inspection,
either to be not rigorously defined, completely impractical, or to violate
criteria such as diachronic consistency that allegedly define 
the very essence of rationality.  
The last two claims --- impracticality and violation of rational
essentials -- surely require mathematical underpinning and 
justification, if they are to have any possible relevance 
to what is presented as a rigorous
mathematical argument.   For example, an account of practicality  
needs some complexity criteria for rational agent
computations: one could then at least discuss the 
empirical justification for the proposed criteria and 
whether and when they actually distinguish
mean utilitarianism from other strategies.  
Similarly, if one accepts that diachronic consistency is 
generically violated and sometimes grossly
violated in the real world, an account of its role in decisions
needs to quantify and compare the degree of violation implied
by different strategies in different circumstances.  
At present, though, the arguments for diachronic consistency and
those concerning practicality rest only
on very debatable verbal intuitions.

As for lack of rigorous definition, there seems to be a danger of a double 
standard, whereby the fuzziness of the ontology is 
used to point out difficulties for alternative strategies (though
in fact it also causes difficulties for mean utilitarianism), while the 
arguments for mean utilitarianism are justified in the context of toy models 
in which a precise definition of a branching structure can be 
found (in which case many strategies other than mean utilitarianism
can be precisely defined).       
The case has not been made that mean utilitarianism is 
well-defined or practical in Wallace's fuzzy Everettian ontology, in which
the mean utility of a strategy can at best only be fuzzily defined.
One can imagine examples in which it either fails to be finite 
or is impractical to estimate -- and it 
seems hard to exclude the possibility that these features often
apply in the real world.  One can also easily 
construct examples in which other strategies are
easier either to approximate or to implement precisely.  
 
Wallace's rationality postulates, likewise, are hard to motivate
in Wallace's fuzzy Everettian ontology, where
they are ultimately intended to apply, but where they are 
difficult, perhaps impossible, to define precisely.  
They generally appear, in any case, possible but uncompelling
guides for rational agents.   Where defined and practical, mean 
utilitarianism is certainly a rationally defensible strategy, with 
some mathematically convenient properties.  But, like 
other critics\cite{pricevol, albertvol}, I am far from persuaded 
that, if I were an Everettian, I should or would be a Born-weighted mean utilitarian.  

\section{Against subjective uncertainty}

One of the stranger claims in the recent Everettian literature is the
suggestion, first made by Saunders \cite{saundersprob, 
saundersvol, saunderswallace, wallaceepiq}, 
that Everettian quantum
theory, although deterministic, nonetheless has a natural
probabilistic interpretation that can be found not by amending the theory or
by adding further postulates, but simply by -- somehow -- analysing the
experience and linguistic usages of agents, that is, creatures 
like ourselves, in an
Everettian universe.  In support of this claim are offered highly
technical and controversial arguments concerning the philosophy of language.
It seems to me simply a mistake, an exercise in wish fulfilment, to 
think that anything of significance to fundamental physics could turn on such 
questions, as though waving the magic wand of linguistic philosophy over
a unitarily evolving state vector could somehow conjure up 
a probability measure and a sample space.\footnote{Incidentally,
this issue has also divided Everettians: Papineau\cite{papineausu} 
and, at one point, Greaves \cite{greavessu},
have also argued that subjective uncertainty is not to be had in 
Everettian quantum theory.}   

Consider Wallace's succinct summary\cite{wallaceepiq} of the argument: 

\begin{quotation}
``[The argument for subjective uncertainty] may be summarised as
follows: in ordinary, non-branching situations, the fact that I expect to become
my future self supervenes on the fact that my future self has 
the right causal and
structural relations to my current self so as to count as my future self. What,
then, should I expect when I have two or more such future selves? There are
only three possibilities:
\begin{enumerate}
\item I should expect 
abnormality: some experience which is unlike normal human
experience (for instance, I might expect somehow to become both
future selves).
\item I should expect to become one or the other future self.
\item I should expect nothing: that is, oblivion.
\end{enumerate}

Of these, (3) seems absurd: the existence of either future self would guarantee
my future existence, so how can the existence of more such selves be treated
as death? (1) is at least coherent -- we could imagine some telepathic link
between the two selves. However, on any remotely materialist account of the
mind this link will have to supervene on some physical interaction between the
two copies -- an interaction which is not in fact present. This leaves (2) as the
only option, and in the absence of some strong criterion as to which copy to
regard as ``really'' me, I will have to treat 
the question of which future self I
become as (subjectively) indeterministic.''
\end{quotation}

This is a false trichotomy.  Consider an (obviously simplified) 
Everettian description of an experiment in which an agent Alice, initially
in brain state $\ket{0}_A$, observes a system in a quantum 
superposition $\sum_{i=1}^2 c_i \ket{i}_S$, where $\ket{1}_S$ and $\ket{2}_S$
correspond, say, to the up and down states of a spin $1/2$ particle,
and becomes entangled: 
\begin{equation}\label{spinexpt}
\ket{0}_A  \sum_{i=1}^2 c_i \ket{i}_S \rightarrow 
\sum_{i=1}^2 c_i \ket{i}_A \ket{i}_S \, .
\end{equation}
Here $\ket{i}_A$ is Alice's brain state after observing the system
in state $i$, for $i = 1,2$.     

Now, as an aside, we actually {\it should}
take possibility (3) seriously, for two reasons. 
First, our conclusions ought to be based on empirical evidence rather
than prejudice.  We 
do not know that Everettian quantum theory is actually correct; we do 
not have a good theory of how consciousness is attached to quantum 
states; we do not know that we or any other agents have ever been 
in a superposition of macroscopically distinct brain states. 
We thus do not know whether, if we were able to place an agent in such
a superposition, they would experience anything --- nor, if so, what.  
Second, there's a coherent view of Everettian
quantum theory in which we are continually being replaced by multiple
copies of future selves.
On this view, even if we assume that superposed selves have individual
experiences, {\it we} will experience 
nothing in future (though our various future selves will).  

The more immediately pertinent point, though, is that 
if we do take Everettian quantum theory seriously, it says, 
indeed, that Alice becomes entangled in a macroscopic superposition.
A coherent way of describing this, which respects the link between 
brain states and mind states, is that just as materially she becomes 
several future selves, her mind becomes several {\it disjoint, non-interacting} 
future minds, with no telepathic link: i.e. option (1) without Wallace's 
misleading gloss.   

The one description that seems obviously wrong, given the rules of the 
game Wallace sets out, is option (2): this really {\it is} an account of 
mind that supervenes on something not present in the physics, namely a 
probabilistic evolution law taking brain state $\ket{0}_A$ to one of the
states $\ket{i}_A$.   

The dangers of attaching some fuzzy 
theory of experience to Everettian quantum theory 
provoke two comments:

First, the fact that we don't 
have a good theory of mind, even in classical physics, doesn't give 
us a free pass to conclude anything we please.   That way lies 
scientific ruin: {\it any} physical theory is consistent with
{\it any} observations if we can bridge any discrepancy 
by tacking on arbitrary assumptions about the link between mind 
states and physics.   We should, rather, be 
all the more cautious and tentative in offering any conclusion.  

Second, the fact that at present no theory of mind can be expressed {\it purely}
mathematically doesn't remove the obligation to strive to express
one's ideas in mathematics {\it as far as possible}.   
Adorning Everettian quantum theory 
with extra assumptions expressed in words -- for instance, as arguments in
linguistic philosophy -- without equations doesn't
alter the fact that one's making extra assumptions: it merely makes 
them more vaguely expressed.  

Consider\footnote{My remarks here follow
Albert's lucid discussion\cite{albertvol}.}
Saunders' exposition\cite{saundersvol}:

\begin{quotation}
Consider a simple example. Alice, we suppose, is about
to perform a Stern-Gerlach experiment; she understands the structure of the
apparatus and the state preparation device, and she is convinced EQM is true.
In what sense does she learn, post-branching, something new? The answer is
that each Alice, post-branching, learns something new (or is in a position to
learn something new) -- each will say something (namely, ``I know the outcome
is spin-up (respectively, spin-down), and not spin-down 
(respectively, spin-up)'') that Alice prior to branching cannot say. 
It is true that Alice, prior to branching, 
knows that this is what each successor will say -- but 
still she herself cannot speak
in this way.
The implication of this line of thought is that, appearances notwithstanding,
prior to branching Alice does not know everything there is to know. What is it
she does not know? I say ``appearances notwithstanding'' for of course in one
sense (we may suppose) Alice does know everything there is to know: she knows
(we might as well assume) the entire corpus of impersonal, scientific knowledge.
But what that does not tell her is just which person she is -- 
or where she is located -- in the wave-function of the universe.
\end{quotation}

But equation (\ref{spinexpt}) suggests there is
no meaning to this question before the 
experiment.\footnote{The same is true in more realistic
models.}
Nothing in the mathematics corresponds to ``Alice, 
who will see spin up'' or ``Alice, who will
see spin down''.  On the left we have  
``Alice, before the experiment''; on the right we have 
``Alice, who has seen spin up'' and ``Alice, who has seen spin down''. 
If one wants to postulate an ``Alice, who will see spin up'', well, 
one can -- but one should then include her in 
the mathematics.
One could, for instance, start with a postulate of the form:

\begin{quotation}
``(P) the probability that A's mind ends up believing that spin is up 
is $| c_0 |^2$ and the probability that A's mind ends up believing 
that spin is down is $ | c_1 |^2$.''
\end{quotation}

This -- Albert and Loewer's ``single mind 
view''  \cite{almanyminds} --
gives only one sentient future Alice.  
To introduce a collection of present Alices who in future
will experience each of the different possible experimental 
outcome, one could, instead, follow Albert and Loewer
in postulating a continuum of Alice minds of which a proportion
$| c_0 |^2$ will see spin up.  One could, in short, adopt the many-minds
interpretation.   I am not persuaded that there is a legitimate alternative 
formulation of Saunders' account.

\appendix

\section{Further comments on counting descendants}\label{countdesc}

Wallace places great stress on the
fact that there is no unique natural definition of a quasiclassical
branch in an Everett universe, and so no way of counting the number of
branches with any given feature.  For example, a naive analysis of a quantum
experiment with three possible outcomes might suggest that a single
branch, pre-experiment, divides into three, post-experiment.  But
this, Wallace stresses, neglects the fact that quantum interactions
take place very frequently in time and densely in space, outside our
control.  A careful attempt to quantify quasiclassical branches would
show many branches splitting into many more during the lifetime of the
experiment; however, there is no unique natural definition that would
allow us to pin these numbers down.  Hence, it is argued, there is no
way of implementing the naive idea of using branch counts to define a
rational strategy -- an approach which would, if it worked, be a
coherent alternative to the Born-rule-dependent strategy, and so
refute the claim that the latter is the unique rational strategy.

There is indeed no known natural way of characterising and counting
branches.  It is worth reconsidering, however, whether there may
nonetheless be a natural way for an agent about to observe an
experiment to characterise and hence count his descendants, by
considering their memory states after the experiment.

Consider first a simple model of 
an observer, apparatus and quantum state which evolve
unitarily during an experiment so that \begin{equation} | O_0 > | A_0
> | Q_0 > \rightarrow \sum_{j=1}^3 a_j | O_j > | A_j > | Q_j > \, ,
\end{equation} where the first state is at time $0$ and the second at
time $t$, the state $ | A_j > $ is the apparatus state registering
outcome $j$ and $ | O_j > $ is the observer state having observed the
apparatus registering outcome $j$.  The agent in state $ | O_0 >$ can
reasonably say that he will have three successors $ | O_j >$ at time
$t$, corresponding to his three future distinct brain states.

Now consider a more detailed model in which we include an environment,
and suppose that 
\begin{equation}\label{modeltwo} | O_0 > | A_0 > |
Q_0 > | E_0 > \rightarrow \sum_{j=1}^3 \sum_{i=1}^{n_j} 
a_{ij} | O_j > | A_{ij} > | Q_{j}> | E_{ij} >  \, .
\end{equation} 
Again, the sum over $j$ represents the three possible
observer states.  The sums over $i$ represent
decompositions into orthogonal quasiclassical branches, and the number
of terms $n_j$ in each sum depends on an arbitrary choice 
of definition of quasiclassical branch from among many possible 
definitions.  Since this is supposed to be a model of
the same experiment, we have that $ \sum_i | a_{ij} |^2 = | a_j |^2 $
for $j = 1,2,3$.  The agent in state $| O_0 >$ has no unique natural
way of characterising the branches.  However, he could consistently view
each of the three components, containing the state $| O_j >$, $j=1,2,3$,
as representing precisely one successor.\footnote{
He is not rationally compelled to accept this view. 
The point is that in this model successor counting is mathematically 
well-defined and hence it's {\it possible} to use it to define
a rational strategy.}  
After all, he is interested in his successors' welfare, and this is
determined by their mind states; at this point in
the analysis, at least, it is not affected by the state of the rest of the
universe.  

It might be objected that we still have not taken sufficiently into
account the pervasiveness of environment-induced quantum interactions,
which will presumably also be taking place within the agent's brain during the
experiment.  A more detailed model still would replace the sums on the
right hand side of (\ref{modeltwo}) by sums including at least small
components of a variety of different agent mind states, corresponding
to different brain states that arise through zapping by stray cosmic
rays and other quantum effects.

A related objection is that this way of counting successors leads to
ambiguities when sequences of experiments are carried out.  Suppose
that the agent will carry out a second experiment, with two possible
outcomes, if he observes outcome $1$ in the first experiment, but not
otherwise.  After the first experiment but before the second, it
appears that he has three successors, whose welfare he should value
equally.  After the second experiment, it appears that he has four
successors, whose welfare he should again value equally.  But since
two of these descend from one of the original successors, and each
``inherit'' any resources he ``bequeaths'' to that successor, the two ways
of counting lead to different allocation strategies -- i.e. they
disagree on which ``bets'' he should be willing to accept on the
experiment.

To these objections, however, the agent could make several responses.

First, that his policy is not incoherent, but merely so far
incompletely specified.  In the case of the two experiments, he does
care equally about the welfare of his three successors between the
experiments, and he also cares equally about the welfare of his four
successors after the experiments.  To formulate a more precise policy,
he would need to set out some way of trading off the welfares of
different successors at different times: for example, by summing the
time-integrated welfares over each distinct life-time segment.  That
this may become complicated doesn't imply that the aim is not
rational.  Indeed, we face very similar problems in worrying about the
well-being of our genetic descendants.  It is perfectly rational to
value the welfare of all of your children equally, and also perfectly
rational to value the welfare of all of your as yet unborn
grandchildren equally, in both cases ceteris paribus.  However,
finding a rational asset allocation policy that respects both these
preferences may require some further policy decisions, complicated
calculations and predictions of uncertain future events.

Second, that he would indeed take into account the welfare of all his
successors, including those whose mind states differ because of
environmentally induced interactions, if he could.  Here again, he can
maintain that he has a rational policy in principle, albeit one that
he cannot fully implement it in practice because of the impracticality
of carrying out the relevant calculations.  And again, he can say that
the latter caveat does not detract from the rationality of his goal.

Third, that a consistent strategy for weighting his concern for the 
welfare of successors could be defined by considering the branching 
structure recorded in their own memory states.   In the example
above, successors who
experience two experiments in succession remember that fact, and 
this distinguishes them from successors who experience only one
experiment.  It's logically consistent -- and not obviously any
more absurd than any Born-weight-dependent many-worlds strategy -- to assign
the former caring weight $1/6$ and the latter caring weight $1/3$. 

To these responses, Everettians might in turn object
that there is no natural way,
even in principle, of characterising and counting all the possible
mind states of successors of an agent exposed to real-world
environmental interactions.  But if the Everettian case eventually
turns on {\it this} point, then 
the objection to rational strategies based on counting successors
ultimately arises from an intrinsic vagueness in the quasifunctionalist 
theory of mind attached to the quantum formalism by Wallace et al.,
not, as claimed, from the vagueness in the notion of a quasiclassical
branch.  It seems a most uncomfortable defence of a
purportedly fundamental theory to say that it is not well enough developed
for us to be able to assess whether or not one of the key arguments
advanced in its favour is valid.  

\section{Further Comment on physical laws of rational 
compulsion}\label{rationalconstraint}

What could it possibly mean to 
believe that the {\it laws of physics per se rationally compel} 
a particular behaviour for rational agents in a branching multiverse?   
The idea here, to be clear, is not merely the truism that the laws of physics 
imply significant facts about the world
which rational agents might, or even must, sensibly take into account. 
It is that there are basic postulates, on an equal footing
with other physical laws, that state by fiat that a particular type of
behaviour is rationally compulsory for rational agents.   
I don't think I know what this can mean -- 
the idea of such a law isn't consistent with my understanding of 
either physics or rationality -- but the 
idea is definitely in play in 
some discussions of many-worlds theories in this volume.  
Papineau\cite{papineauvol} 
proposes an axiom of this type, and Greaves-Myrvold\cite{greavesmyrvoldvol}
consider how to (purportedly) confirm theories including such 
axioms.   My impression is that many
Everettians share Greaves' view\cite{greaves} that resorting to such a 
postulate would be, at least, an adequate fall-back 
should Wallace's\cite{wallacevoltwo} and other
arguments not hold up.   

Here's a point that seems not to have been 
considered in the Everettian literature.   
If we {\it were} to take seriously the
idea that physical axioms can rationally compel rational beings to act in 
a certain way -- by fiat, without further justification -- then
we must also take seriously the possibility that these rationally
compelling axioms can take unfamiliar forms.   For instance, 
in our branching universe, there's no reason to restrict to 
axioms that require rational preferences to be 
given by the ordering of values of expressions of the form
$ \sum_i p_i U_i $, where $U_i$ is the agent's utility for 
the outcome on branch $i$ and the $p_i$ are positive branch weights 
satisfying the Kolmogorov axioms. 
There's nothing {\it logically} inconsistent about postulating
laws with negative or complex $p_i$, or preference orderings given by
general joint functions $ f( { p_i } , { U_i } )$, or indeed any other
mathematical structure one cares to dream up.\footnote{Indeed, one 
could imagine an   
exotic story about inverted qualia and hence reversing of utilities
on some branches, which justifies giving them negative weights.   
And perhaps, some might argue, the complex quantum amplitudes defining the path
integral should be interpreted as directly defining rational constraints
on an agent's preferences for the entire set of paths defining a hypothetical
future unitary evolution.}

Such laws would generally violate 
Savage's axioms and perhaps other cherished intuitions about rational 
behaviour.   But once one enters the strange game of postulating
{\it physical} laws {\it defining} rational behaviour in 
multiverses, one needn't restrict one's postulates
to intuitions developed in an attempt to provide a foundation 
for decision theory in a single chancy universe.   
Everettians who miss this point seem to me like 
hypothetical seventeenth century theorists who learn Hooke's law, 
come up with the idea that one can postulate abstract physical force laws that 
define forces between objects unmediated by springs, but then still maintain
that these laws necessarily have to set force proportional to 
separation.   Their boldness is inconsistently selective: an
abstract law need not be constrained  
by the details of the concrete model that inspired it.   

Of course, Everettians who think it makes sense to  
postulate laws of rational compulsion still have the option of  
basing their postulates on one-world probability theory, and
specifically on optimising Born-rule-weighted average utility.  
But one needs to be clear that, 
{\it prima facie}, this is an arbitrary choice from a very 
large range of possibilities.   Along with everything else in this 
peculiar game, that choice seems to lack justification.
Moreover, as the above analysis of Greaves-Myrvold's account of confirmation applied 
to the weightless universe shows, allowing 
arbitrary choices of laws of rational compulsion means not only that 
many mutually inconsistent choices can be postulated in the same
multiverse but also that each of them can be, by their own lights, confirmed. 

\section{A possible empirical distinction between many-worlds and one-world
quantum theory}

Finally, suppose, notwithstanding all the arguments above, 
that we arrive at an Everettian theory that, while perhaps 
ad hoc and unattractive, is coherent -- for example, some version of 
the many-minds interpretation\cite{almanyminds}.  
It is generally believed that, without very advanced technology which
allows the re-interference of macroscopically distinct branches, 
such a theory will necessarily be empirically indistinguishable from Copenhagen 
quantum theory.  

The following argument against this conclusion relies on anthropic 
reasoning and also on the hypothesis that species may evolve a  
consistent preference for or against higher population expectation over higher 
survival probability.  Anthropic reasoning is notoriously 
tricky to justify, and we may anyway not necessarily have evolved 
demonstrable consistent preferences one way or the other, so 
the argument may not necessarily have practical application.   
Nonetheless, it does
show in principle that evolutionary evidence could make many-worlds 
theories more or less plausible. 

Consider a simple model of two species $A$ and $B$, both of which begin
with population $P$ and are
offered, each year, the option of doing something that depends on a quantum
event and carries a $0.5$ probability of extinction and a $0.5$ probability 
of trebling the species population.  Suppose that, if they reject
the option, their 
population remains constant,
as it does in between these decisions.  Species $A$ is risk-averse, and so
always declines the option.   Species $B$ is risk-tolerant, and 
instinctively driven to maximise expected population, and so always
accepts.  

Now let $N$ be a large integer.  
After $N$ years, if one-world quantum theory is correct, species $A$ 
will have population $P$, and species B will have either population
$0$ (with probability $ ( 1 - ( \frac{1}{2} )^N )$) or population
$3^N$ (with probability $ ( \frac{1}{2} )^N$).   In other words, species
$B$ will almost surely be extinct.   If these are the only two species,
and you are alive in the $N$-th year, almost certainly you belong to 
species $A$.   

If many-worlds quantum theory is correct, species $A$ still has 
population $P$ in all branches.   Species $B$ has population $0$
in branches of total Born weight $ ( 1 - ( \frac{1}{2} )^N )$, and 
population $3^N$ in branches of total Born weight $ ( \frac{1}{2} )^N$. 
Now, if anthropic reasoning is justifiable here, and you are alive in 
the $N$-th year, almost certainly you belong to species $B$.
(There are $( \frac{3}{2} )^N$ times as many minds belonging to species
$B$ as to $A$ after $N$ years.)  

In other words, there is a sense in which long-run evolutionary success 
is defined by different measures in one-world and many-worlds quantum theory.
If anthropic reasoning were justifiable, then one could in principle infer 
whether one-world or many-worlds quantum theory is likelier correct
by seeing whether one belongs to a Born-weighted expected
population maximising species or to a risk-averse species that seeks
to maximise its Born-weighted survival probability.  
Readers may thus wish to consider whether their species 
has evolved a coherent strategy of either type.\footnote{Unfortunately,
I suspect mine has not.}

\acknowledgments  I am very grateful to Jonathan Barrett
for many thoughtful and constructive comments on and criticisms of 
a preliminary version of the manuscript, as well as other very
valuable conversations, and also to Hilary Greaves and David Wallace 
for patiently tolerating and helpfully engaging with my critical probing.  
Thanks too to David Albert, Harvey Brown, Jeremy Butterfield, Chris Fuchs,
Lucien Hardy, Graeme Mitchison, Wayne Myrvold, David Papineau, Huw Price, 
Simon Saunders, Tony Short, John Sipe, Rob Spekkens and Tony Sudbery for 
some valuable conversations. 
This research was partially supported by a grant from The Foundational
Questions Institute (fqxi.org) and by 
Perimeter Institute for Theoretical Physics.  
Research at Perimeter Institute is supported by the Government of 
Canada through Industry Canada and by the Province of Ontario 
through the Ministry of Research and Innovation.

\end{document}